Title: Assessing the Global Wind Atlas and local measurements for bias correction of wind power generation simulated from MERRA-2 in Brazil

Authors: Katharina Gruber[a], Claude Klöckl[a], Peter Regner[a], Johann Baumgartner[a], Johannes Schmidt[a]

[a]Institute of Sustainable Economic Development, University of Natural Resources and Life Sciences, Feistmantelstraße 4, Vienna, Austria

katharina.gruber@boku.ac.at, claude.kloeckl@boku.ac.at, peter.regner@boku.ac.at, johann.baumgartner@boku.ac.at,

Corresponding author: Katharina Gruber

# Assessing the Global Wind Atlas and local measurements for bias correction of wind power generation simulated from MERRA-2 in Brazil

**Abstract**

NASA's MERRA-2 reanalysis is a widely used dataset in renewable energy resource modelling. The Global Wind Atlas (GWA) has been used to bias-correct MERRA-2 data before. There is, however, a lack of an analysis of the performance of MERRA-2 with bias correction from GWA on different spatial levels – and for regions outside of Europe, China or the United States. This study therefore evaluates different methods for wind power simulation on four spatial resolution levels from wind park to national level in Brazil. In particular, spatial interpolation methods and spatial as well as spatiotemporal wind speed bias correction using local wind speed measurements and mean wind speeds from the GWA are assessed. By validating the resulting timeseries against observed generation it is assessed at which spatial levels the different methods improve results – and whether global information derived from the GWA can compete with locally measured wind speed data as a source of bias correction. Results show that (i) bias correction with the GWA improves results on state, sub-system, and national-level, but not on wind park level, that (ii) the GWA improves results comparably to local measurements, and that (iii) complex spatial interpolation methods do not contribute in improving quality of the simulation.

**Keywords:** wind power simulation, bias correction, Brazil, Global Wind Atlas, MERRA-2 reanalysis

**Nomenclature**

| Symbol | Meaning |
|---|---|
| **Data** | |
| GWA | Global Wind Atlas, mean wind speeds retrieved from the Global Wind Atlas at 50 m height |
| IN | INMET - Instituto Nacional de Meteorologia, wind speeds retrieved from the National Meteorological Institute of Brazil at 10 m height |
| MER | MERRA-2 - Modern-Era Retrospective analysis for Research and Applications, Version 2; wind speeds at 10 m and 50 m height |
| ONS | Operador Nacional do Sistema Elétrico, National Electrical System Operator of Brazil, capacities and historical power generation are taken from this source |
| TWP | The Wind Power, source for wind park data (locations, installed capacities, commissioning dates) |
| **Symbols in formulae** | |
| $w$ | Wind speed at one location |
| $w_{h,d,m,y}^{MER10}$ | MERRA-2 wind speeds at 10 m height at hourly resolution at one location |
| $w_{h,d,m,y}^{MER50}$ | MERRA-2 wind speeds at 50 m height at hourly resolution at one location |
| $w_{h,d,m,y}^{MER108}$ | MERRA-2 wind speeds interpolated to 108 m height at hourly resolution at one location |
| $w^{GWA}$ | Global Wind Atlas mean wind speeds at 50 m height at one location |
| $w_{h,d,m,y}^{IN}$ | INMET wind speeds at hourly resolution at 10 m height at one measurement station |
| $w_{h,d,m,y}^{c}$ | Wind speeds corrected by mean wind speed interpolated to 108 m height at hourly resolution at one location |
| $w_{h,d,m,y}^{chm}$ | Wind speeds corrected by hourly and monthly correction factors and interpolated to 108 m height at hourly resolution at one location |
| $cf\_cap$ | Correction factor for capacity |
| $cap\_ons$ | Capacities of the National Electrical System Operator of Brazil per day |
| $cap\_twp$ | Capacities in The Wind Power dataset per day |
| **Indices** | |
| h | Hour |
| d | Day |
| m | Month |
| y | Year |
| **Results** | |
| MBE | Mean bias error |
| RMSE | Root mean square error |
| BLI | Bilinear Interpolation |
| IDW | Inverse Distance Weighting |
| NN | Nearest Neighbour method |
| wsma | Wind speed mean approximation (using GWA) |
| wsc_hm | Wind speed correction with hourly and monthly correction factors |

# 1    Introduction

NASA's MERRA and MERRA-2 reanalysis data sets are commonly used sources of climate data for simulating long timeseries of renewable energy generation, and in particular wind power generation (see Table 1). While they have been applied successfully in many contexts, the MERRA datasets have two disadvantages: they are well known to show significant deviation from observed means and their spatial resolution is rather low at 50 km. One way of making it spatially more accurate, is horizontal interpolation to the location of simulated wind turbines. Additionally, the Global Wind Atlas (GWA) was applied to MERRA data in a few recent studies [1, 2, 3, 4] to decrease the bias in the dataset.

Considering spatial interpolation methods, a majority of studies apply bilinear interpolation for obtaining wind speeds at particular locations from the gridded reanalysis data, some use LOESS interpolation [5, 6], while others [1, 7] mention that they apply interpolation methods but do not further specify which (see Table 1). However, none of the studies made a systematic effort to understand the impact of different methods of spatial interpolation on the quality of simulations, which is why the authors see the need to study this in detail.

With regard to bias-correction, there are some examples where it is applied to increase the quality of simulation results (see Table 1). However, these sources rely on historical wind power generation data, which comes with two disadvantages. First, these data need to be available for each country where simulation is desired and second, this harbours the risk of overfitting. A more convenient way is to use the GWA, which provides two major benefits with its comparably high resolution and global availability. Nevertheless, it has only been applied in a few cases: Gonzalez-Aparicio et al. [1] compared the performance of MERRA and ECMWF reanalysis and their EMHIRES dataset downscaled with GWA data for the simulation of European wind power generation. Two other studies by Bosch et al. similarly apply the GWA for spatial downscaling of MERRA-2 wind speeds in order to simulate global onshore [3] and offshore [2] wind power potential, while also considering the factors of land-use, topography and technology. Ryberg et al. [4] also apply the GWA, as well as air density for bias correction in a study on the European wind energy potential and the impact of turbine design on wind power simulation. However, none of them assessed how different data sources used for bias-correction affect the quality of simulation. The authors therefore conduct such an analysis to fill this gap of knowledge.

From previous work, it can also be identified that most of the research is geographically restricted to one particular or a selection of European countries, and usually considers nationally aggregated power generation (see Table 1). Only one study was identified to study a region in the US [8], whereas assessments of Chinese wind power potentials [9, 10] as well as global potentials [2, 3] do not validate their simulated time series. Furthermore, merely three studies could be identified, apart from aggregated wind power output, to also validate wind power generation at particular wind parks in Sweden [8] as well as the Netherlands and France [4], whereas Drew et al. [11] validate wind power generation for a cluster of offshore wind farms.

We conclude that existing studies show, in principle, that using reanalysis data for simulation of wind energy generation gives reasonable results on aggregated levels, a thorough assessment of the quality on different spatial levels in combination with different methods for spatial interpolation and for data sources of bias correction is however lacking currently – as well as studies outside of Europe and the US.

The authors aim at answering the following research questions: (1) How much do interpolation methods contribute to the accuracy of wind power generation simulated from reanalysis data? (2) Can the Global Wind Atlas compete with local measurements in mean bias correction? (3) Can results be further improved using seasonal and diurnal bias correction? (4) How does spatial aggregation affect the quality of simulation? The present study contributes here by thoroughly assessing the quality of wind power simulations from MERRA-2 data, applying and comparing spatial interpolation methods and different bias-correction methods and data sets for the case of Brazil. Brazil, is in size, comparable to Europe, has a growing wind power fleet [12], attaining 14.3 GW in 2018 [13], and the available generation data allows an assessment of simulation quality on the level of wind parks, states, sub-systems and the complete national system. Besides contributing to a better understanding of how to best use MERRA-2 data for simulation of wind energy generation, the produced timeseries are of high interest for the long-term planning of Brazil's energy system.

Table 1: Summary of previous studies and their scopes, as well as the datasets, interpolation methods, bias correction and validation they use

| Authors | Ref. | Topic | Dataset | Interpolation method | Bias correction | Validation | Area |
|---|---|---|---|---|---|---|---|
| González-Aparicio et al. | [1] | Simulating European wind power generation, statistical downscaling | MERRA, ECMWF, GWA | | Wind (GWA) | Wind, wind power (1 year) | Europe |
| Bosch et al. | [2] | Offshore wind energy potentials | MERRA-2, GWA | Bilinear | Wind (GWA) | None | Global |
| Bosch et al. | [3] | Onshore wind energy potentials | MERRA-2, GWA | | Wind (GWA) | None | Global |
| Ryberg et al. | [4] | European onshore wind energy potential and turbine design | MERRA, GWA | Bilinear | Wind (GWA), air density | Wind power (wind parks in NL and FR, country DK) | Europe |
| Staffell and Pfenninger | [5] | Bias correction of current and future wind power simulation | MERRA, MERRA-2 | LOESS | Wind power | Wind power | 23 European countries |
| Staffell and Green | [6] | Decline of wind farm performance with age | MERRA | LOESS | Load factor (weather) | Wind, wind power | UK |
| Monforti and González-Aparicio | [7] | Uncertainties of technical and meteorological parameters in wind power simulation | MERRA | | | None | Europe |
| Olauson | [8] | Comparison of MERRA-2 and ERA5 | MERRA-2, ERA5 | Bilinear | Wind power (1/2 years sample data, rest testing) | Wind power | Wind farm level: Sweden, country level: Germany, Denmark, France, Sweden, Bonneville Power Administration |
| Ren et al. | [9] | Characterization of wind resource | MERRA-2 | | | Wind | China |
| Ren et al. | [10] | Spatio-temporal complementarity of renewable energy | MERRA-2 | | | None | China |
| Drew et al. | [11] | Impact of offshore wind farms | MERRA | Bilinear | | Wind power (cluster of offshore wind farms) | Great Britain |
| Cannon et al. | [14] | Quantifying extreme wind power generation statistics | MERRA | Bilinear | | Wind, wind power | Great Britain |
| Olauson and Bergkvist | [15] | Modelling Swedish wind power production | MERRA | Bilinear | Wind power | Wind power | Sweden |
| Kubik et al. | [16] | Sensitivity analysis of wind power simulation | MERRA | Bilinear | | Wind power | Northern Ireland |
| Cradden et al. | [17] | Simulation of wind generation and impact of pressure patterns | MERRA | Bilinear | | Wind, wind power | Ireland |
| Olauson and Bergkvist | [18] | Correlation of wind power generation between European countries | MERRA | | Wind power | Wind power | Europe |
| Kubik et al. | [19] | Regional wind power simulation | MERRA | | | Wind power | Northern Ireland |
| Johansson et al. | [20] | Assessing marginal system value of wind turbines | MERRA, ERA-Interim | | Wind (ERA-Interim mean) | None | Europe |
| Huber et al. | [21] | Integration of wind and solar power in Europe | unpublished dataset based on MERRA | | | Wind power (ramps, Germany and Ireland) | Europe |

## 2 Data & Methodology

The graph in Figure 1 gives an overview of the data and methods used for simulating and bias-correcting wind power generation as well as the data used for validation and subsequent analysis. The method can be described in three steps:

1. Wind power generation was simulated while testing four simple interpolation methods with little computational effort for reanalysis wind speeds: Nearest Neighbour, Bilinear and Bicubic Interpolation and Inverse Distance Weighting. Resulting time series were compared to observed wind power generation by statistical evaluation and the best method was selected.

2. Two different sources of wind speeds were used for bias correcting the mean wind speeds: wind speed measurement data from the national institute of meteorology of Brazil (INMET) and mean wind speeds from DTU's Global Wind Atlas (GWA). The results of the correction with both sources were validated against wind power generation data and the better one was determined.
3. The final step consisted in adding the temporal component to the spatial bias correction with hourly INMET data, where it was tested whether hourly and monthly wind speed correction improved the fit of wind power generation to observed data.

In this study, several sources of data were used for the purpose of generating a model, which simulates wind power generation output from reanalysis wind speed data in Brazil. The used datasets are listed in Table 2, together with their temporal availability (at the moment of download). In the Appendix additional information on installed capacity per investigated region and the beginning of time series are specified. The complete source code can be found on github [22].

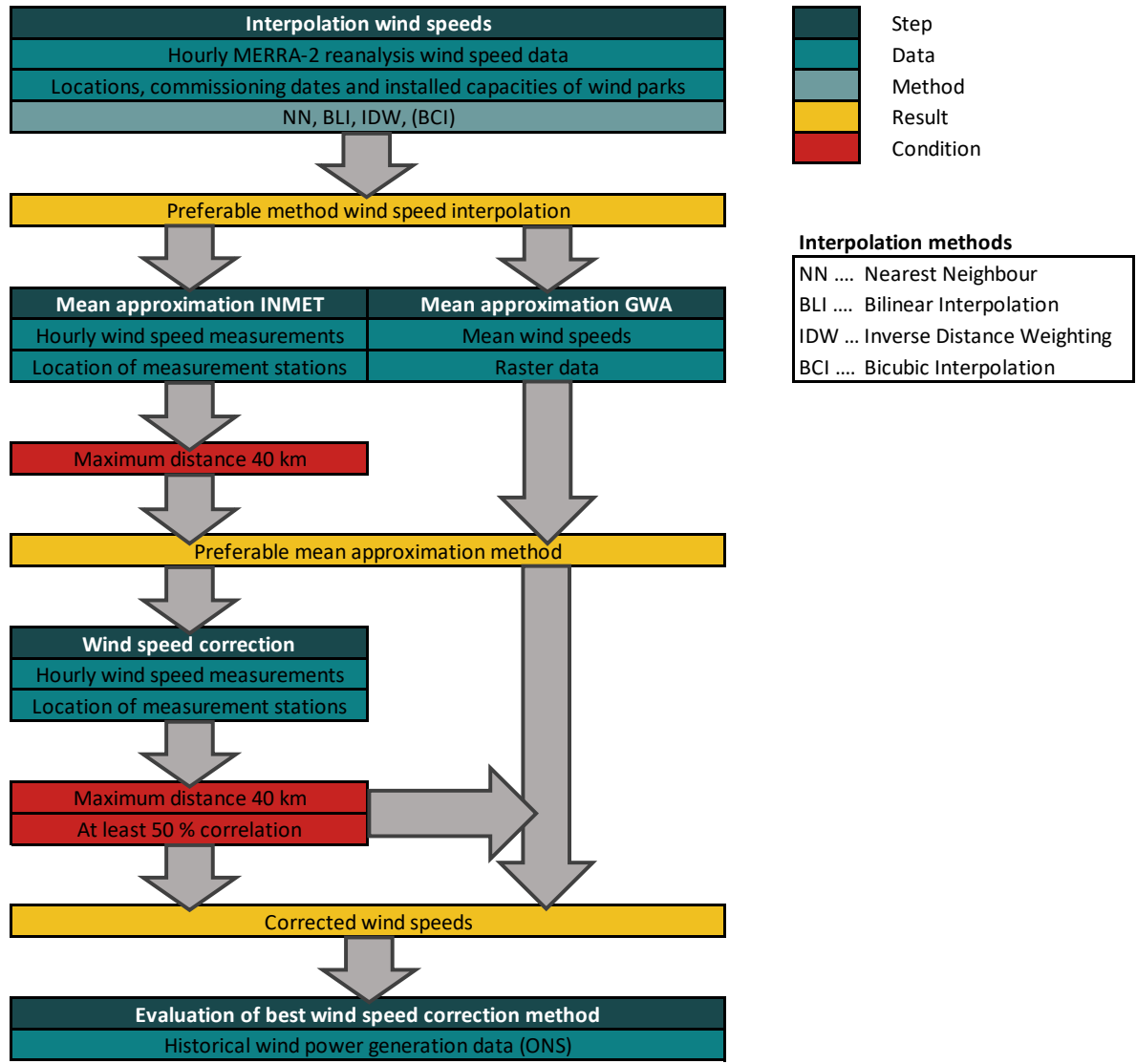

*Figure 1: Overview of the approach: used data and methods for the wind power simulation model and bias correction*

*Table 2: Summary of data used for modelling of wind power and for analysis*

| Data set name | Description | Temporal resolution | Coverage | Source |
|---|---|---|---|---|
| **MERRA-2** | Reanalysis data, modelled wind speed data | Hourly | 1980-Aug 2017 | NASA |
| **BDMEP** | Wind speed measurement data | Hourly | 1999-2016 | INMET |
| **Global Wind Atlas** | Mean wind speeds | Mean | 2015 | DTU, IRENA |
| **Wind farms** | Wind park data, geographical locations and installed capacities with commissioning dates (complemented with data from different sources) | Monthly | 1998-2017 | The Wind Power |
| **Histórico da operação** | Historical wind power generation data | Daily | 2006-Oct 2017 | ONS |

### 2.1 Simulation of wind power generation

For the simulation of wind power generation locations, capacities as well as commissioning dates of present wind parks were required. This information was retrieved from The Wind Power website and comprises the name of the wind farm, the country and county (state) it is located in, the municipality at which it is located, the commissioning date, the number and type of installed wind turbines, the installed capacity, and the geographical coordinates. A few wind parks were lacking information (installed capacities, geographical location, commissioning date, state) which was complemented from other sources. For example, information from the National Agency of Electrical Energy of Brazil (ANEEL) [23] which provides wind parks with the municipality they are located in was used. Figure 2 depicts location, commissioning year, and capacity of Brazilian wind parks and shows the main wind power generation regions: the North-East and South of Brazil (also see A.1 ). Wind power generation was calculated starting in 2006, as prior to that no noteworthy capacities were installed (eight wind parks have commissioning dates before 2006 with a capacity of 28.1 MW in total) and there are no data for comparison available for the period before 2006. Furthermore, the graph shows that the majority of wind power plants were installed in the past eight years.

The MERRA-2 (Modern-Era Retrospective analysis for Research and Applications, Version 2) data which were used as a source of wind speed data are a reanalysis dataset provided for free by the National Aeronautics and Space Administration (NASA) [24]. Wind speed data in u- and v- direction at three different heights as well as the according displacement height are available in temporal resolution of one hour and spatial resolution of about 50 km between data points (0.625° longitude and 0.5° latitude). Data are available since 1980 and updated monthly.

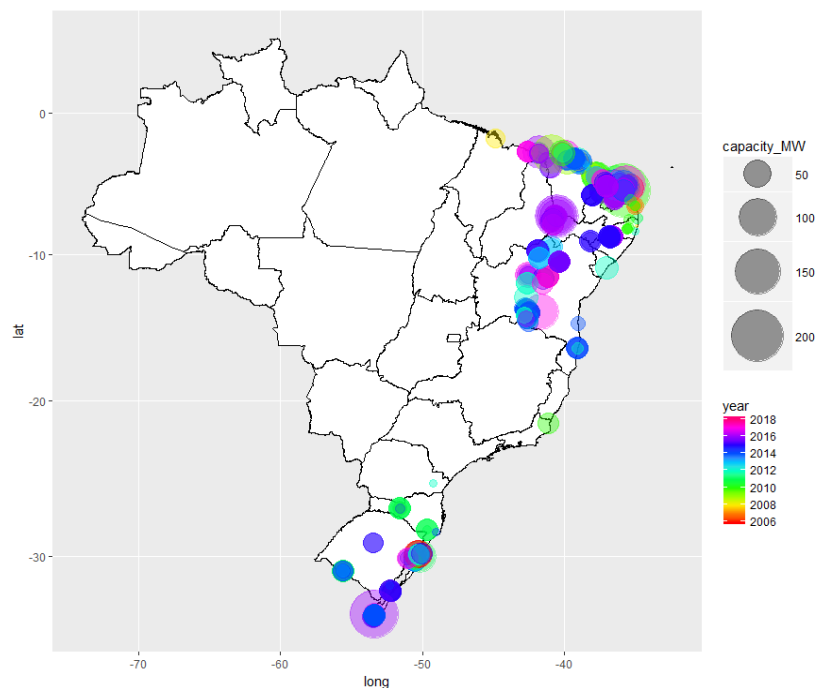

*Figure 2: Location, capacity and commissioning year of wind parks in Brazil*

For determining wind speeds at specific locations from reanalysis data, four different methods were tested: the Nearest Neighbour Method, the Bilinear Interpolation, the Bicubic Interpolation, and Inverse Distance Weighting (see Table 3). The Nearest Neighbour method is the simplest of these methods and consists in using data from the grid point which has the smallest geographical distance to the desired point of interpolation [25]. It is not only fast, but also suitable if many data points are available [26]. Bilinear interpolation uses four surrounding points and is a simple, yet accurate method, which can be applied if data are available in a grid [27]. Inverse Distance Weighting is a method where it is assumed that data from surrounding points around the point of interest influence data on this point inversely to their distance. It is fast and can be used for interpolation of irregularly distributed data [25]. For keeping this method simple and fast and also because grid points are not very close with a distance of about 50 km, only four neighbouring points were considered for interpolation. Inverse Distance Weighting is a commonly used method in meteorology [26], which in several cases has been proved to be the best method for interpolation among others [28] [29], such as Kriging. As Bicubic Interpolation delivered negative wind speeds in some cases it was discarded for further use.

*Table 3: Summary of interpolation methods with short description*

| Method | Abbr. | Description |
|---|---|---|
| **Nearest Neighbour Method** | NN | Wind speeds of the closest MERRA-2 grid point were used |
| **Bilinear Interpolation** | BLI | The four surrounding points in a square around the point of interest were first linearly interpolated in one direction (resulting in two points), which were afterwards also interpolated in the other direction |
| **Bicubic Interpolation (not applied)** | BCI | 16 closest points in square around point of interest were inserted in a cubic equation and with the resulting coefficients calculated by solving the equation system, the wind speed at the point of interest was determined |
| **Inverse Distance Weighting** | IDW | Wind speeds were calculated on the point of interest from the four closest grid points, where each of these points was weighted inversely to its distance from the point of interest |

Effective wind speeds were calculated by the Euclidean norm from wind speeds in u- and v-direction. For inter- and extrapolation of wind speeds to certain heights (of reference wind speeds as well as of wind turbines) the wind profile power law was applied, where the wind speed in a certain height depends on the ratio of the heights and an exponent (alpha friction coefficient), which is higher the lower the surface roughness. The alpha friction coefficient was determined from the wind speeds in two different heights (10 m above disposition height and 50 m above ground), which were chosen due to their proximity to the heights used in the simulation. Different approaches were also tried but discarded due to implausible or worse results (see A.4 ).

According to results of a previous analysis, the use of specific details of installed turbines in the simulation instead of using one generic power curve will only marginally affect results. An analysis carried out as part of the present work, however, showed, that applying site specific turbines can improve results, compared to using standard turbines. Therefore, for simulation of wind power from wind speeds the power curve model developed by Ryberg et al. [4] was applied, taking into account turbine specific information. The model is based on estimating the power curve of a wind turbine from the specific power of a wind turbine. After simulating a single turbine, the capacity was scaled to the installed power at the level of the wind park. The specific power needed as input parameter for the power curve model was not given in the wind park data, but can be calculated by division of turbine capacity by rotor area. In most cases information on the turbine capacity was included in the dataset. The rotor diameter, which is needed to calculate rotor area, was extracted from the wind turbine dataset, where available. It was also attempted to obtain a more accurate model by including information on the hub height. As this information was not available in the wind turbine dataset, a linear regression model was fitted to the turbines in the US Wind Turbine Database [30], such that hub heights could be estimated from rotor diameters (Figure 3). Where not available, specific power and hub height were estimated from the mean specific power or hub height of turbines installed in the same year.

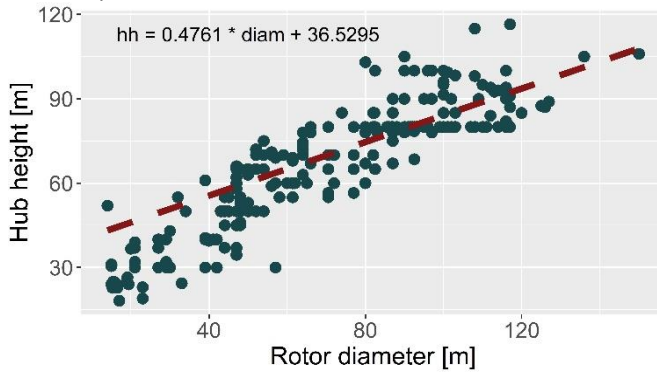

*Figure 3: Linear regression of hub height (hh) on rotor diameter (diam) based on data from the US Wind Turbine Database [30]*

**2.2 Bias correction**

As mentioned before, reanalysis data can have significant bias. In particular wind speeds in MERRA-2 are subject to bias due to the rather coarse spatial resolution of the underlying model, which is why the present study aimed at reducing this bias with various correction methods. For this purpose, reference wind speed data from two different sources were compared: Wind speed measurement data provided by the National Meteorological Institute of Brazil (Instituto Nacional de Meteorologia, INMET[1]) as well as mean wind speeds in the Global Wind Atlas (GWA) [31] from DTU's Department of Wind Energy (Danmarks Tekniske Universitet[2]) available on the IRENA (International Renewable Energy Agency[3]) website. INMET data are available since 1999 in hourly resolution at 10 m height above ground for 481 wind speed measurement stations, of which 478 are on Brazilian mainland (see Figure 4). The GWA mean wind speeds are provided in a raster with 1 km x 1 km resolution in three heights (50 m, 100 m and 200 m above surface). The high resolution of the GWA was harnessed to increase the spatial detail of the coarser reanalysis dataset without applying methods of high computational effort.

For applying bias correction, the reference wind speed ($w^{MER}_{h,d,m,y}$) point closest to the GWA point or INMET measurement station was identified and its data was used for mean approximation or temporal wind speed correction. Mean approximation was performed by multiplying with the proportion of mean observed ($w^{GWA}$ or $mean(w^{IN}_{h,d,m,y})$) and mean reanalysis ($mean(w^{MER50}_{h,d,m,y})$ or $mean(w^{MER10}_{h,d,m,y})$) wind speeds (Eq. 1). In order to apply INMET wind speeds for correction (Eq. 2), $w^{GWA}$ was replaced by $mean(w^{IN}_{h,d,m,y})$ and $mean(w^{MER50}_{h,d,m,y})$ was replaced by $mean(w^{MER10}_{h,d,m,y})$), both either at 50 m above ground for bias correction with the GWA or at 10 m above ground for the bias correction with INMET data; temporal wind speed correction by aggregating wind speeds in specific hours and months (Eq. 3, $\sum_{d,y} w^{IN}_{h,d,m,y}$ and $\sum_{d,y} w^{MER10}_{h,d,m,y}$) and multiplying reanalysis wind speeds with the according proportions. Temporal wind speed correction was only possible with INMET data, as GWA data does not contain temporally resolved information.

$$w^{c}_{h,d,m,y} = w^{MER108}_{h,d,m,y} * \frac{w^{GWA}}{mean(w^{MER50}_{h,d,m,y})} \quad (1)$$

$$w^{c}_{h,d,m,y} = w^{MER108}_{h,d,m,y} * \frac{mean(w^{IN}_{h,d,m,y})}{mean(w^{MER10}_{h,d,m,y})} \quad (2)$$

$$w^{chm}_{h,d,m,y} = w^{MER108}_{h,d,m,y} * \frac{\sum_{d,y} w^{IN}_{h,d,m,y}}{\sum_{d,y} w^{MER10}_{h,d,m,y}} \quad (3)$$

While the GWA is a globally, spatially continuously available dataset that does not need any further treatment, the INMET dataset is based on stationary data, which are not available on a regular grid and where data quality issues remain. The dataset therefore had to be cleaned and rules on which stations to use for bias-correction needed to be implemented.

First, data cleaning was performed on measured wind speeds: an analysis of the time series revealed some long sequences of the same wind speed in measured data, especially unusually long sequences of 0 m/s wind speeds. This was considered to be an error in the data. These erroneous sequences were removed using a threshold of five days, meaning any row of same values with a length of at least 120 hours, is discarded. Also missing values (NAs) were removed from measured data before comparison of the time series.

As wind speed measurement data from INMET are only available at specific locations, wind speed correction is not possible for wind parks where the distance to the closest INMET data point is too high. A constraint of 40 km maximum distance to the closest INMET wind speed measurement station was introduced. Tests have shown that this distance leads to the smallest errors (see A.5). Furthermore, it is about the maximum distance of the closest MERRA-2 point to a location, restricting correction to INMET station on the same MERRA-2 tile. Figure 4 shows which wind parks were not bias corrected with INMET data due to this restriction.

A second restriction was implemented for the hourly and monthly wind speed bias correction regarding the correlation between wind speed time series after wind speed bias correction. In some cases, wind speed time series and reanalysis data showed very low correlations. As measurements can be erroneous or local conditions may differ significantly compared to those on the closest reanalysis data point, the authors introduced a threshold in which case mean approximation is applied instead of temporal bias correction. A minimum of 50 % correlation after correction is required for application of hourly and monthly wind speed correction.

---

[1] http://www.inmet.gov.br/
[2] http://www.dtu.dk/
[3] http://www.irena.org/

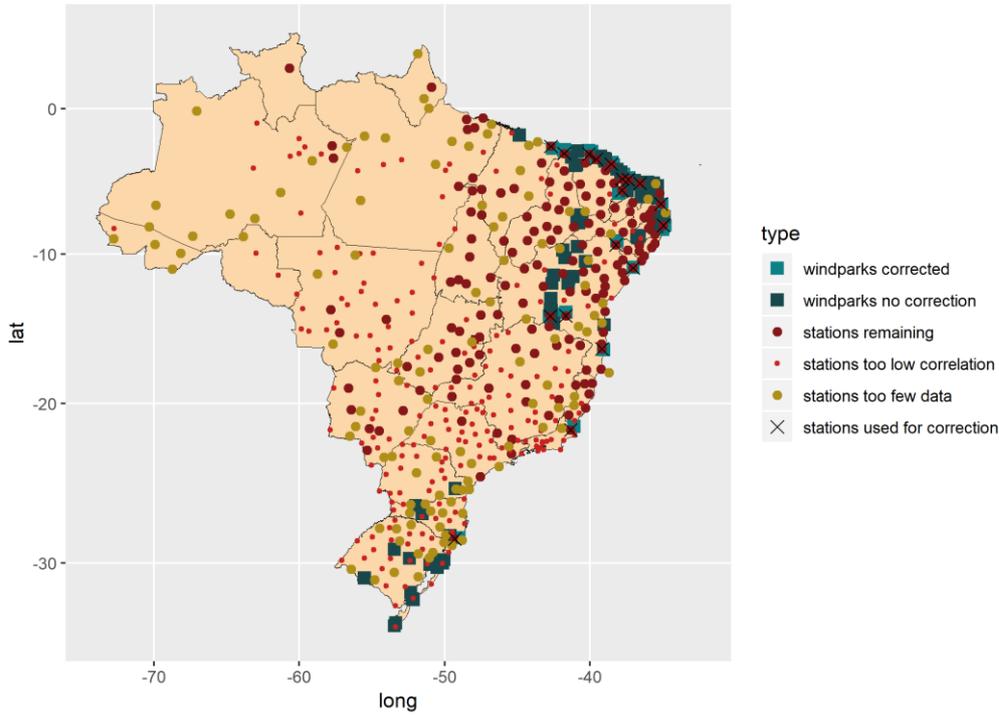

*Figure 4: Location of INMET wind speed measurement stations and wind parks (depiction with data from [32], [33] and [34])*

A third restriction was applied on whether to use the data of a specific wind measurement station depending on data quality and availability. For this purpose, three limits were set: In the years since 1999 at least four complete datasets had to be available for each month, i.e. in at least four years there had to be data for 30 days of January, March, April etc. (with the exemption of February). If these conditions were satisfied for a particular station, only data of months which provide at least ten days (240 hours) of data were used. Months with less than ten days of available data were excluded from the analysis. When these restrictions were implemented, only 365 of the 478 wind speed measurement stations were qualified for hourly and monthly wind speed correction or mean approximation with INMET data – not considering distance or correlation limits. Figure 4 shows the locations of INMET wind speed measurement stations (red and gold) and wind parks (blue and dark blue) using wind speed bias correction with Nearest Neighbour Interpolation. Wind speed measurement stations available for wind speed correction (i.e. with proper data quality, availability and sufficient correlation between observed and reanalysis wind speeds) are presented in red, those actually utilised for correction are highlighted with Xs. Wind speed measurement stations discarded for their low correlation are represented by smaller light red points and those with insufficient data by gold points. Wind parks subject to wind speed correction are marked with blue squares, those that remain uncorrected with dark blue squares.

During a preliminary analysis some significant differences between simulated and observed wind power generation were found. Discrepancies in installed capacities used for simulation and reported by the ONS were examined as one possible reason. From Brazil's national system operator monthly time series of installed wind power capacities are available for the whole country or per subsystem. On these levels, mean capacities from ONS and The Wind Power were compared, and capacity bias correction factors $cf\_cap$ (Eq.4) for Brazil, the North-East and the South were determined for the whole period. These are all below 1 (Brazil 0.93, North-East 0.92, South 0.99), meaning that the capacities given by The Wind Power are decreased to the level of ONS capacities.

$$cf\_cap = \frac{mean(cap\_ons)}{mean(cap\_twp)} \qquad (4)$$

**2.3 Validation and selection of methods**

Wind power generation data for validation of simulated wind power time series is available on the homepage of the electrical grid operator (Operador Nacional do Sistema Elétrico, ONS) of Brazil [35]. These data can be downloaded in different spatial and temporal resolutions: for the whole of Brazil, for three Brazilian subsystems, for eight states or even for single wind parks time series of daily, weekly, monthly or yearly wind power generation are available for download. For this study, daily data were applied on all spatial disaggregation levels. Wind power generation currently occurs in the subsystems South, North-East and North. However, the latter, which only has one wind power generating state (Maranhão - note that, although Maranhão belongs geographically to the North-East, in ONS it is classified in the North subsystem) was not considered separately in the analysis (but added to Brazilian production), as wind power generation is comparatively low and started only recently (May 2017) there, limiting the means for comparison. On the level of wind parks, in each state preferably the wind park with the largest installed capacity was selected for comparison of wind power generation time series, provided data of observed and simulated wind power generation were available. In some cases, names of wind parks in the wind parks dataset and in the ONS database did not always match or were not available from ONS. The wind parks selected for the analysis are listed in Table 4.

The wind power generation time series from these wind parks as well as for the larger areas were assessed with simple error measures: in the first step of method selection, i.e. the choice of interpolation methods, correlations were most relevant and chosen as criterion for selection. In the second step, i.e. the selection of a data source for wind speed mean approximation, the focus was on reducing the average bias between the simulation and observed wind power generation, which was represented by the root mean square error (RMSE) and the mean bias error (MBE) for different levels of spatial disaggregation. For the third step, i.e. the temporal bias correction, also the correlation was of interest, as it was tested whether it could be increased by applying temporal bias correction. In the Appendix some additional measures are provided.

*Table 4: Locations selected for the analysis of simulation results on the level of wind parks*

| Windpark | State | Comment |
|---|---|---|
| **Macaúbas** | Bahia | Third largest, the two largest are new and time spans for comparison are short |
| **Praia Formosa** | Ceará | Largest |
| **São Clemente** | Pernambuco | Consists of 8 parts, only matching wind park in Pernambuco, only short time series of 1.2 years available |
| **Araripe III** | Piaui | Only matching wind park, less than one year of data available |
| **Alegria II** | Rio Grande do Norte | Second largest wind park, after São Miguel do Gostoso, which has no available data |
| **Elebras Cidreira 1** | Rio Grande do Sul | Second largest, for largest wind park (Hermenegildo) no historical generation available |
| **Bom Jardim** | Santa Catarina | Only matching wind park |

## 3    Results

In this section the main outcomes of the analysis of the simulation of wind power generation in Brazil with different interpolation and bias correction methods will be presented. Additional results can be found in section A.2. The results of the three steps (i) simulation with interpolated reanalysis data, (ii) bias correction of the mean wind speed, and (iii) temporal wind speed correction are analysed and a method selection is performed to limit the amount of results to relevant ones. Results are displayed as graphs for easier comparison. A collection of statistical parameters can be found in the Appendix (Table A 2, Table A 3, Table A 4).

The correlations of simulated and observed daily wind power generation shown in Figure 5 are usually high, at least at spatially aggregated levels (Brazil and subsystems), where they are above 94 %. Using Bilinear Interpolation or Inverse Distance Weighting instead of the Nearest Neighbour method does not have any significant impact. For lower spatial aggregation levels, i.e. states and particular wind parks, correlations are different from the higher aggregated levels, but no differences in the quality of interpolation methods can be observed. Correlations are in a similar range as for Brazil and its subsystems in most cases, except in Pernambuco, where it is significantly lower - but this applies to all examined interpolation methods. For particular wind parks, correlations are in general lower, ranging between 0.5 and 0.9, indicating a better quality of simulation at the higher level of spatial aggregation. Overall, no preferable method can be selected from evaluation of correlations, as the differences are only minor.

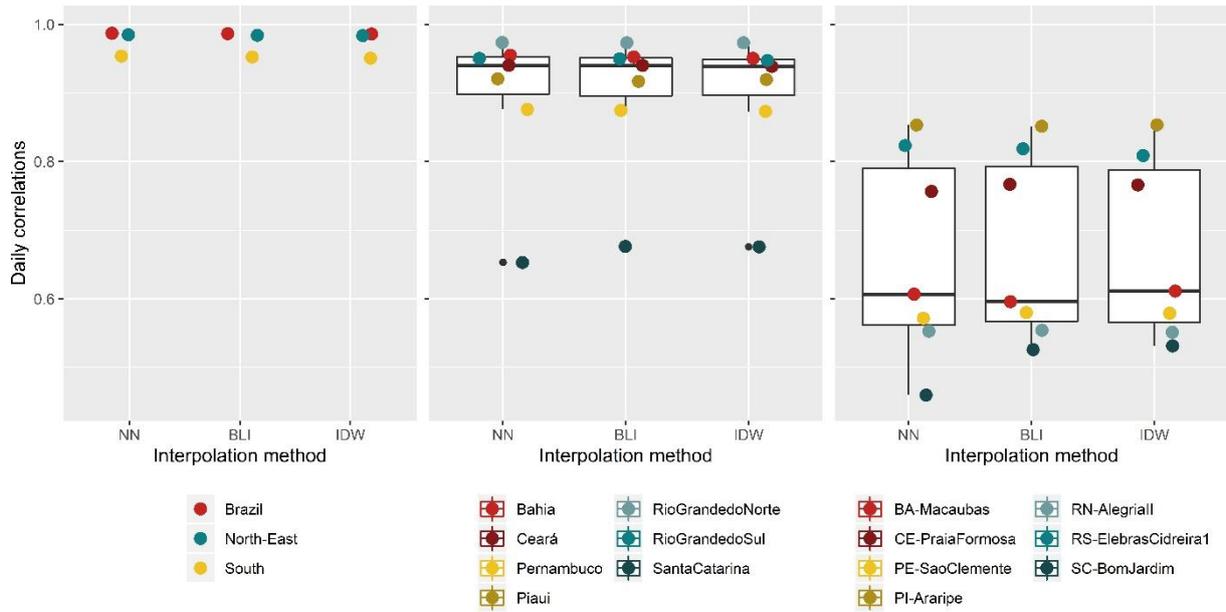

*Figure 5: Correlations of observed and simulated daily wind power generation time series with three interpolation methods (Nearest Neighbour: NN, Bilinear Interpolation: BLI, Inverse Distance Weighting: IDW) for Brazil, the North-East and South subsystems, seven states and seven selected wind parks*

If relative RMSEs (Figure 6) are compared for the tested interpolation methods, no significant differences can be observed either. For certain regions (Brazil, South, Rio Grande do Sul, Alegria II, Araripe) RMSEs are slightly higher when using the Nearest Neighbour method, but for others (Ceará, Praia Formosa) RMSEs are lowest when applying this simple method.

Other statistical parameters and plots used for comparison are shown in the Appendix (Table A 2, Figure A 1). However, neither MBEs nor comparison of means indicate a favourable method. The present results show that the Nearest Neighbour method yields comparably good results in terms of correlations and RMSEs of simulated and observed wind power generation time series in particular for aggregated regions, and thus was chosen as the favourable method, as it is very simple and has low computational needs. For simulating single wind parks, it may be beneficial to test other methods too, as in some cases the results of the Nearest Neighbour method were slightly worse than the other tested approaches.

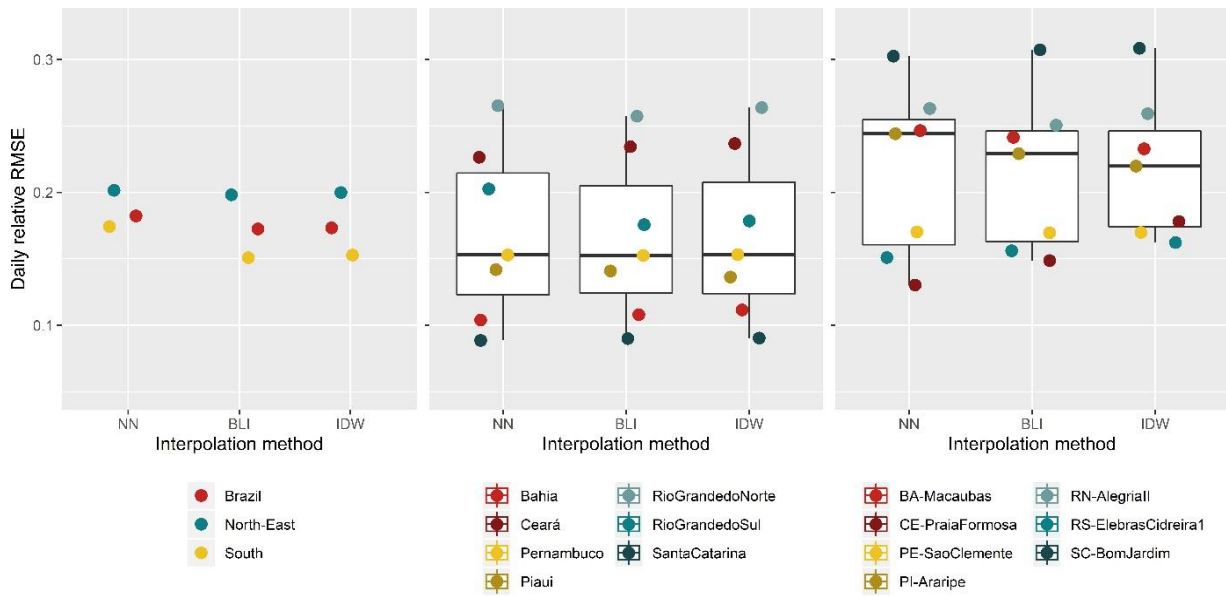

*Figure 6: Relative RMSEs of observed and simulated daily wind power generation time series with three interpolation methods (Nearest Neighbour: NN, Bilinear Interpolation: BLI, Inverse Distance Weighting: IDW) for Brazil, the North-East and South subsystems, seven states and seven selected wind parks*

In order to reduce the bias between observed and simulated wind power generation time series, bias correction of wind speeds was applied. The first step consisted in selecting a favourable data source for mean wind speed correction: The two sources examined in this study were the mean wind speeds from DTU's Global Wind Atlas (GWA) and measured wind speeds from the national meteorological institute of Brazil (INMET). Three measures for bias, the RMSE, the MBE and the deviation in means were examined. Only the relative RMSEs and MBEs (which apart from the bias also indicates whether observed wind power generation is over- or underestimated), normalised by the mean installed capacity in the period of validation for easier comparison between different regions, are shown here. The other parameters can be found in the Appendix (Table A 3, Figure A 2, Figure A 3, Figure A 4). An investigation of this error measure on the level of Brazil or the subsystems (Figure 7) showed a clear tendency: The highest RMSEs occur if no mean approximation is applied, which can be reduced by applying bias correction with GWA mean wind speeds and even more so by applying correction with measured wind speeds (INMET). If looking at spatially disaggregate analysis, results are slightly different: On the level of states, correction with measured wind speeds results in the lowest RMSEs on average. What is striking, however, is that on the level of states and wind parks mean approximation with INMET has strongly negative impacts for selected locations (Pernambuco, Santa Catarina and the single wind parks evaluated in these states). Overall, INMET bias correction seems to deliver the best results as it reduces RMSEs for most locations at spatial aggregation levels above wind parks compared to no bias correction and also if mean approximation with GWA wind speeds is applied. For other statistical parameters, results are slightly different.

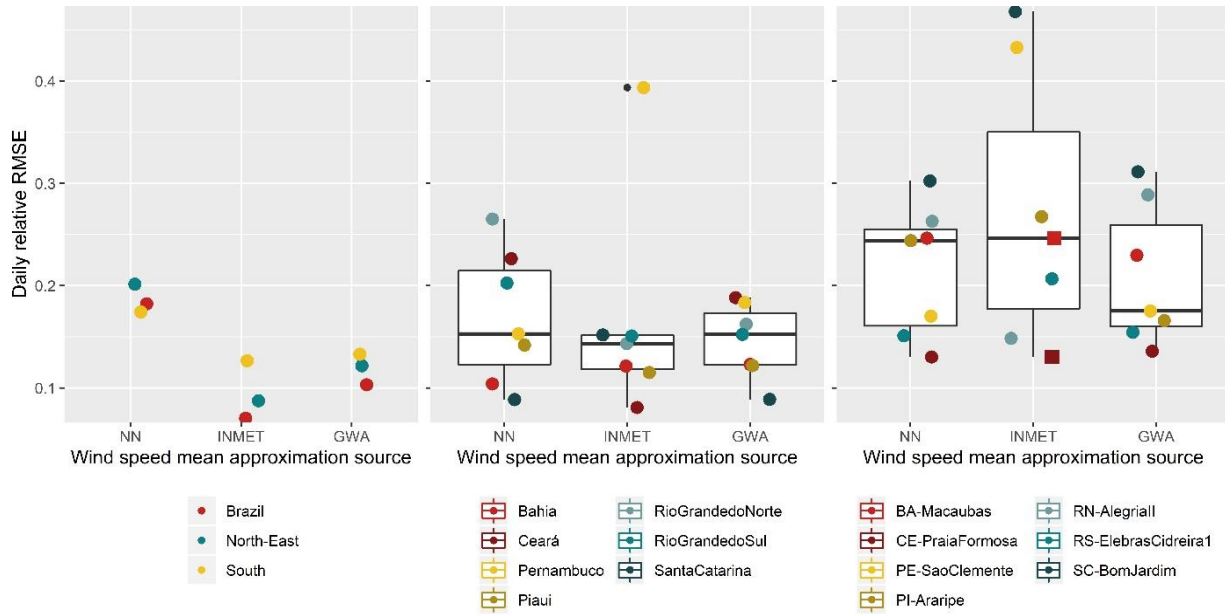

*Figure 7: Comparison of relative root mean square errors (RMSEs) of simulated daily wind power generation with different sources for wind speed mean approximation: INMET wind speed measurements (INMET), Global Wind Atlas mean wind speeds (GWA) and no correction (Nearest Neighbour, NN). The two wind parks marked as squares were not corrected because they are too far away from the closest INMET wind speed measurement station.*

Figure 8 shows the relative mean bias errors (MBEs) between observed and simulated daily wind power generation using different data sources for mean bias correction. According to the MBE indicator, similarly to the results from the RMSEs, it is more recommendable to apply mean approximation with INMET data in Brazil and the North-East subsystem, as the values are closer to 0 than with GWA or without correction. Mean bias correction with INMET data brings lower errors for most states, while increasing the bias for Pernambuco and Santa Catarina considerably. On the level of wind parks, however, results of wind speed mean correction are inconclusive. Correction with measures wind speed data caused benefits only for Alegria II whereas in São Clemente, the error was increased drastically.

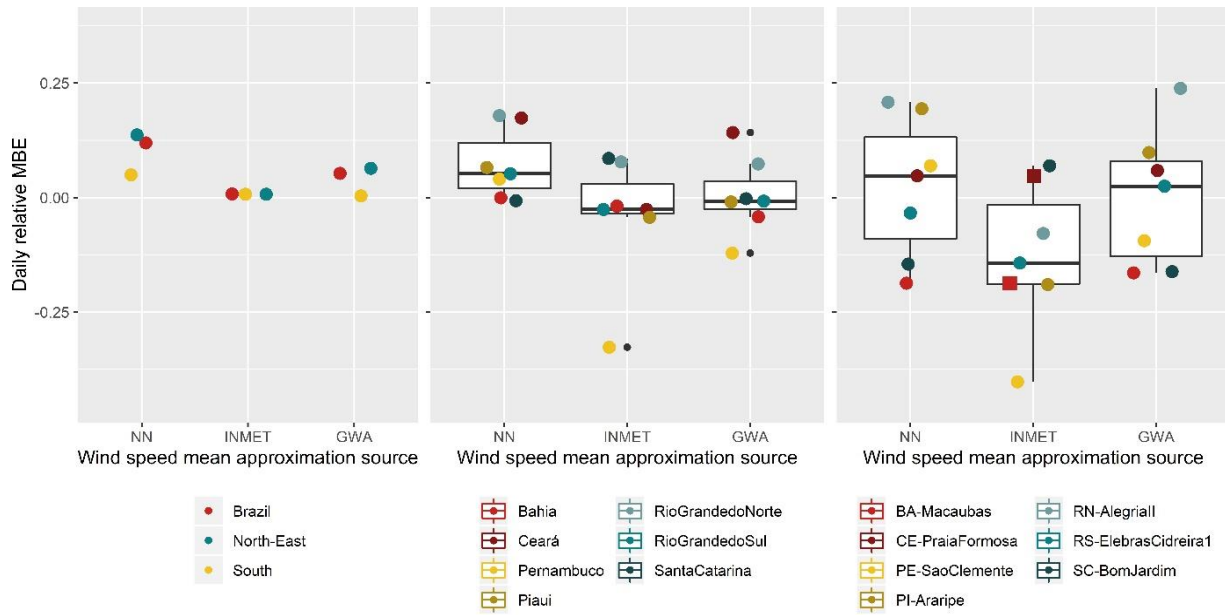

*Figure 8: Comparison of relative mean bias errors (MBEs) of simulated daily wind power generation with different sources for wind speed mean approximation: INMET wind speed measurements (INMET), Global Wind Atlas mean wind speeds (GWA) and no correction (Nearest Neighbour, NN). Note that for Bom Jardim the value is missing for correction with INMET because it is far higher than the other values shown and therefore out of range of the graph. The two wind parks marked as squares were not corrected because they were too far away from the closest INMET wind speed measurement station.*

In fact, it could not be determined definitely which data source is better for wind speed mean approximation, especially as different statistical parameters do not always result in the same conclusions. Results with GWA were not always the best, but rarely the worst. Wind speed mean approximation with measured wind speeds sometimes delivered good results, compared to mean approximation with GWA wind speeds or without wind speed mean correction, for example in Brazil, its North-East, in Ceará or at the wind park in Alegria II. In other cases, however, it led to considerable reduction of wind speeds, resulting in underestimation of wind power generation, as seen in Pernambuco, or the wind parks of Araripe or Sao Clemente. When using GWA wind speeds for correction, however, resulting simulated wind power generation usually was in a similar range as the observed and no extreme outliers were observed. Therefore, GWA was considered to be the more stable source for bias-correction and was used for further analysis.

In the next step, GWA mean approximation was combined with a more precise method of wind speed correction with the help of measured wind speeds. Hourly and monthly wind speed correction was tested, to see if spatiotemporal correction can improve results compared to only spatial bias correction. Figure 9 shows the relative RMSEs between simulated and observed wind power generation time series. Hourly and monthly correction resulted in lower RMSEs in the North-East, but higher in the South. On the level of states, spatiotemporal wind speed correction increased RMSEs for Pernambuco and Rio Grande do Sul, while for the wind park Alegria II the error between simulated and observed wind power generation was decreased substantially. However, spatiotemporal wind speed correction with measured wind speeds sometimes shows a positive impact on correlations for higher levels of aggregation (Table A 4).

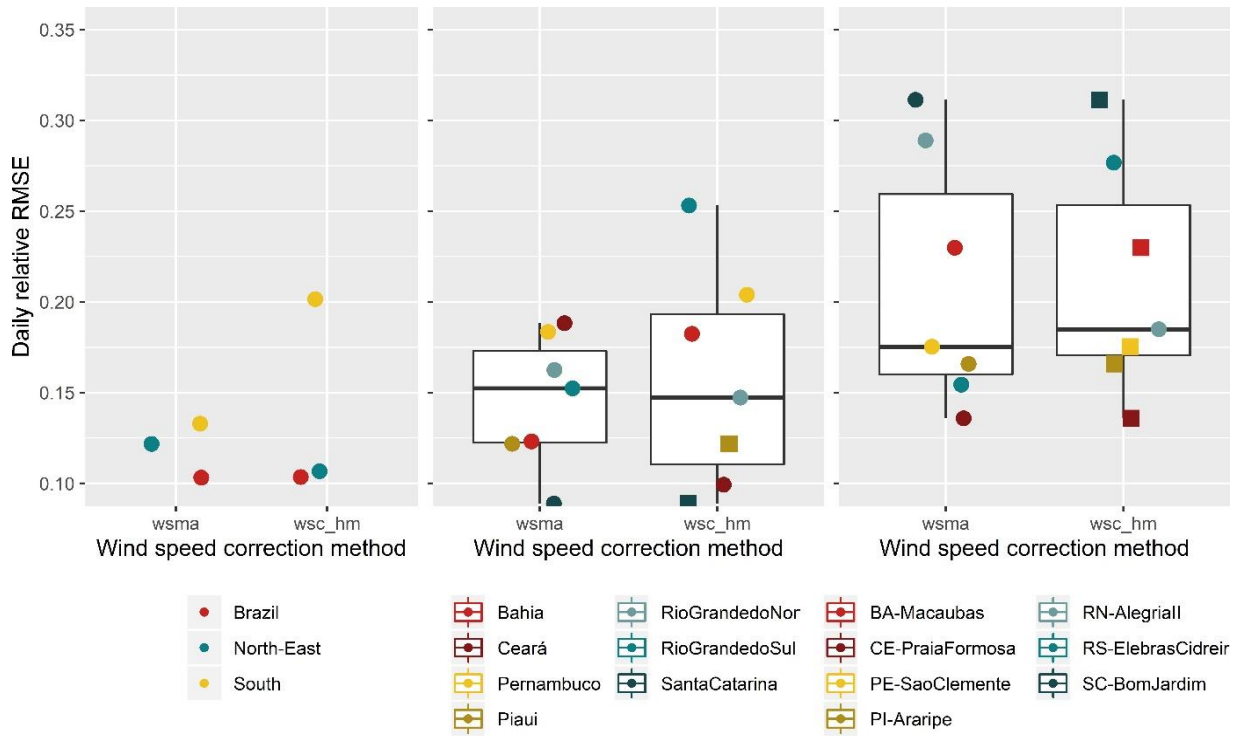

*Figure 9: Comparison of relative root mean square errors (RMSEs) of simulated daily wind power generation with different methods for wind speed bias correction: mean approximation with Global Wind Atlas data (wsma) and mean approximation with Global Wind Atlas data combined with hourly and monthly wind speed correction with INMET wind speed data (wsc_hm). The locations marked as squares are not corrected because they are too far away from the closest INMET wind speed measurement station or correlation after correction is below 50%.*

When assessing the bias by MBEs (Figure 10), results were partly different from those from RMSEs. Similarly to the RMSEs, a reduction in the error can be observed in Brazil and the North-East, but not in the South. On the level of states results are ambiguous. For some a slight shift away from 0 can be observed in the relative MBEs when using the hourly and monthly wind speed correction, while in other states the error is decreased, which fits the observation made from RMSEs. For particular states (except those who are not corrected), only the errors of Rio Grande do Norte and Ceará are decreased, while in Pernambuco and Rio Grande do Norte the MBEs shift away from 0 (also see Table A 4, Figure A 5, Figure A 6 and Figure A 7). For individual wind parks, the overall picture shows a shift away from 0 and a general reduction of MBEs when applying spatiotemporal bias correction, however the bias of the only two wind parks which are corrected was reduced in Alegria II, while in both cases leading to a less realistic underestimation of wind power generation. This shows, that it may be useful to apply this kind of bias correction to some regions in order to reduce the error between simulated wind power generation and observed data, especially on a local scale and in the North-East and Brazil, but can increase the bias in specific states.

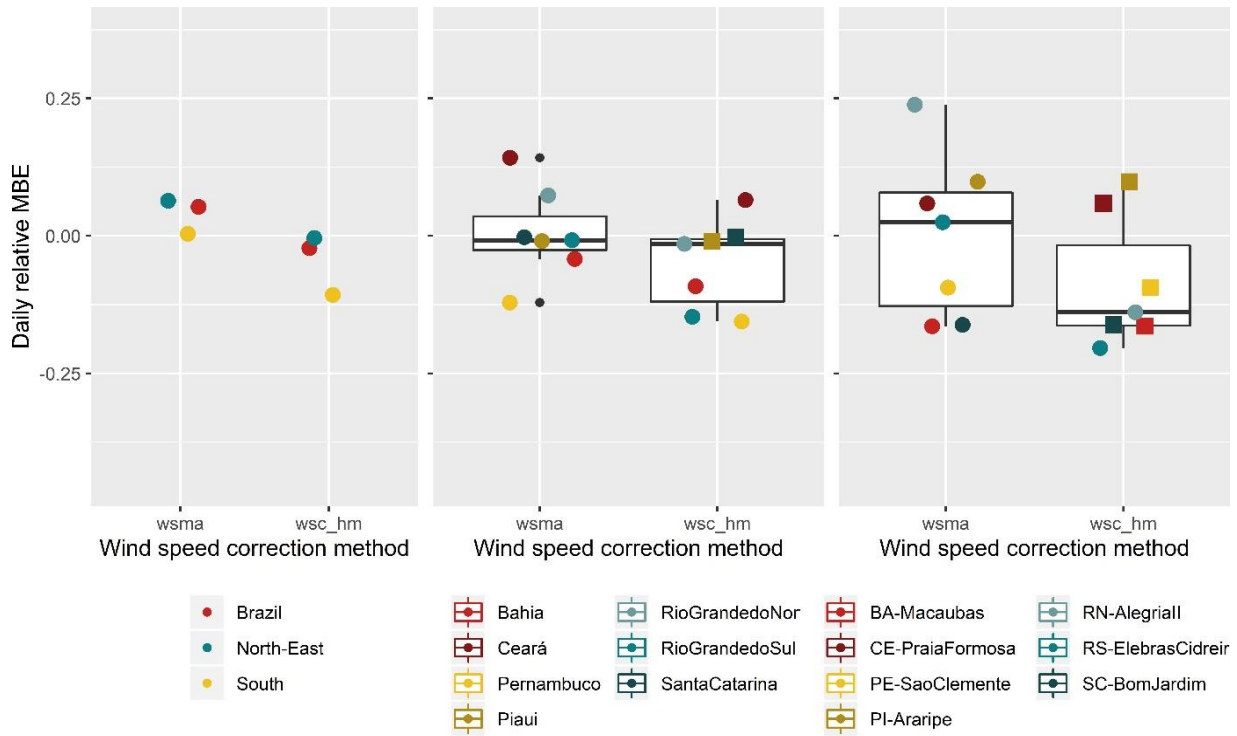

*Figure 10: Comparison of relative mean bias errors (MBEs) of simulated daily wind power generation with different methods for wind speed bias correction: mean approximation with Global Wind Atlas data (wsma) and mean approximation with Global Wind Atlas data combined with hourly and monthly wind speed correction with INMET wind speed data (wsc_hm). The locations marked as squares are not corrected because they are too far away from the closest INMET wind speed measurement station or correlation after correction is below 50%.*

## 4 Discussion

In the first part of the analysis three spatial interpolation methods for gridded wind speeds were tested for the simulation of wind power. This study's results showed that more advanced methods (BLI and IDW) could not contribute to higher correlations compared to the Nearest Neighbour method. Especially if spatially aggregated areas are examined, there was no improvement in correlations. For single wind parks, there may be some benefits in using BLI and IDW.

Considering correlations, the present results are similar to those reported in other studies (see Appendix Table A 5), such as by Cannon et al. [14], Cradden et al. [17], or Pfenninger and Staffell [5]. Only González-Aparicio et al. [1], who study simulation of wind power generation in European countries using three different wind speed data sets, report lower correlations than ours. Evaluating relative biases and RMSEs (see Appendix Table A 5) only González-Aparicio et al. [1] obtained some similar values to those presented here, or few even higher (Sweden and Switzerland), whereas in other cases RMSEs [17, 5] and biases [17] were lower than in the present study.

Part of this can be explained by a more inhomogeneous topography in Brazil compared to countries like Ireland or Germany which are analysed in the other studies, especially due to the large spatial extent. It has to be noted, however, that these examples consider monthly [17, 5] and hourly [1, 14, 5] time series, while in this study daily wind power generation was considered, which can have an impact on results. On the other hand, it should be considered that the area of Brazil is larger than European countries which is likely to have a higher smoothing effect. This shows in the correlations which are high and thus indicate a good simulation quality by MERRA reanalysis data, confirming the Cannon et al. results [14] for a different world region.

González-Aparicio et al. [1] also found that the data for comparison of power time series from the transmission system operators show some inhomogeneities, which may be a possible explanation for some of the error in the present study.

Furthermore, they state that another possible source of error in MERRA data, or wind power generation calculated from those data is that the coarse spatial resolution results in an underestimation in variability of wind speeds, especially in areas of complex terrain. The fact that reanalysis data often neglect local conditions and therefore may lead to some bias is stated by others too, such as Cannon et al. [14], Pfenninger and Staffell [5] or Olauson and Bergkvist [15]. Previous research has shown that such bias can be reduced by spatially aggregating power output of several wind farms, as then the simulation takes advantage of smoothing effects [36, 37].

The second part of the present work focused on reducing the bias between simulated and observed wind power generation by applying wind speed correction, comparing two different sources for reference wind speeds. Results from this section did not indicate a clear tendency whether correction with measured wind speeds or with mean wind speeds delivers a smaller error between simulated and historical wind power generation. However, it showed that bias correction in many cases had a positive effect on the simulation, especially in areas of higher spatial aggregation (Brazil and North-East). There, lowest RMSEs were obtained when wind speeds are corrected with INMET data, but MBEs were closer to 0 when approximating to the GWA. On the levels of states, correction with GWA mean wind speeds mostly led to a better fit of the simulation to observed wind power according to MBEs. A probable conclusion that can be drawn from this result is that spatially detailed information

– i.e. in the present case GWA data - is especially important if wind power generation for smaller areas is modelled because there the relative error is high, however, less important when larger areas are considered. One would also expect a positive impact of adding this information on the level of wind parks. However, the data quality (especially of observed wind power generation data) may not be sufficient to see the benefits of this.

Nevertheless, as recommended also by Monforti and Gonzalez-Aparicio [7], correction of reanalysis data should be applied, at least at larger spatial levels. Otherwise, according to Rose and Apt [38] who examined the variability of wind energy in the U.S. Great Plains by simulation based on reanalysis data, it is likely that reanalysis data might underestimate wind speeds and thus wind power generation for particular locations. This cannot be supported by this study's results, as on the level of wind parks, wind power simulated directly from reanalysis data usually was in a similar range as the historical data, or slightly above. When measured wind speeds were used in bias correction, a higher share of the simulations underestimated observations. This is considered less realistic, as maintenance times and wake losses were not considered in the simulation, which should result in a slight overestimation compared to observed wind power generation. Maintenance times, however, will have a low impact on overall output assuming that they take place during times of low wind power generation. Another study [1] which took a similar approach simulating wind power generation from MERRA data with bias correction with GWA wind speeds, generating a new dataset called EMHIRES, found that adding spatially detailed information improved the representation of historical data of wind power generation. Bosch et al. [2, 3] used a similar approach with GWA data for simulating wind power generation from MERRA-2 wind speeds. They however only calculated wind power potentials for several countries while assuming that GWA provided more accurate data, thus allowing no basis for comparison to the present results.

In the last step of the analysis, bias correction was refined. Measured wind speed data were used to not only correct the overall mean but also seasonal and diurnal means of wind speeds, by applying hourly and monthly wind speed correction factors. However, for the majority of regions and wind parks assessed, the fit of the simulated time series to historical wind power generation data did not improve or results were rather inconclusive. Only three states, as well as the North-East and Brazil gained higher correlations with hourly and monthly wind speed correction. According to the present analysis, the average bias could not consistently be further reduced by temporal bias correction with INMET data, especially as correction was hardly applied due to the set restrictions. Overall, the authors determined that for spatially aggregated areas spatiotemporal bias correction is not necessary, although it could reduce bias in particular cases. For individual locations it can be useful – but only if data in good quality are available. Other studies provided no means of comparison as temporal bias correction, if performed, relied on wind power instead of wind speed data, such as the investigation of Olauson and Bergkvist [15], or directly used measured wind speed data for simulating wind power generation and compared the results to TSO and simulations from reanalysis [19].

## 5     Conclusion

In this paper the quality of simulating wind power generation in Brazil from MERRA-2 reanalysis data was assessed and different approaches for interpolation and bias correction with local and global sources on different spatial levels were compared. The aim was to select the best of the examined methods for generation of wind power time series. Results showed that (i) horizontal interpolation methods of higher computational effort are usually not necessary as they did not improve results, (ii) bias-correction with GWA delivered results comparable to locally measured data and in general improved results compared to simulations without any correction on all spatial levels with the exception of single wind parks, (iii) spatiotemporal bias-correction is only advised if high quality measured data is available and (iv) none of the available datasets or methods could consistently improve simulation results on the level of wind parks, where the error in general was higher compared to spatially aggregated areas, i.e. a mean relative RMSE of 0.21 on the level of wind parks, compared to a relative RMSE of 0.10 on the level of Brazil when using GWA for bias correction.

In the future, results found in this study can be applied to simulate wind power generation time series, which can consequently be used to assess potentials of renewable energy. The outcome that GWA data could contribute to smaller bias in estimation of wind power generation, is especially important as this method can be applied globally, paving the way for studies considering the entire world or at least different spatially distant regions such as Europe and the Americas.

**Acknowledgements**
This project has received funding from the European Research Council (ERC) under the European Union's Horizon 2020 research and innovation programme (grant agreement No. 758149).

**Supplementary Material**
Data are available online. DOI: 10.5281/zenodo.3460291

[32]    G. Cruz, F. Estrela, B. Junior and M. Lima, "Shapefiles do Brasil para download," CodeGeo, 16 Apr 2013. [Online]. Available: http://www.codegeo.com.br/2013/04/shapefiles-do-brasil-para-download.html. [Accessed 05 Jan 2018].

[33]    Brasil Governo Federal, "INMET Instituto Nacional de Meteorologia," [Online]. Available: http://www.inmet.gov.br/portal/. [Accessed 11 Feb 2018].

[34]    M. Pierrot, "The Wind Power. Wind Energy Market Intelligence," The Wind Power, [Online]. Available: https://www.thewindpower.net/. [Accessed 11 February 2018].

[35]    ONS, "GERAÇÃO DE ENERGIA," ONS, 2018. [Online]. Available: http://www.ons.org.br/Paginas/resultados-da-operacao/historico-da-operacao/geracao_energia.aspx. [Accessed 05 Jan 2018].

[36]    R. Goić, J. Krstulović and D. Jakus, "Simulation of aggregate wind farm short-term production variations," Renewable Energy, pp. 2602-2609, 1 May 2010.

[37]    F. J. Santos-Alamillos, D. Pozo-Vázquez, J. A. Ruiz-Arias, V. Lara-Fanego and J. Tovar-Pescador, "A methodology for evaluating the spatial variability of wind energy resources: Application to assess the potential contribution of wind energy to baseload power," Renewable Energy, pp. 147-156, 2014.

[38]    S. Rose and J. Apt, "What can reanalysis data tell us about wind power?," Renewable Energy, pp. 963-969, 05 Jun 2015.

[39]    I. González-Aparicio and A. Zucker, "Impact of wind power uncertainty forecasting on the market integration of wind energy in Spain," Applied Energy, pp. 334-349, 16 Sep 2015.
17

# Appendix
## A.1  Investigated regions and installed capacities
*Table A 1: Mean and maximum installed capacities and start dates of simulation and validation time series of investigated regions*

| Region | Startdate | | Installed capacity [MW] | |
| --- | --- | --- | --- | --- |
| | Simulation | Validation | Mean | Max |
| **Brazil** | 2006-01 | 2006-03 | 3043 | 11749 |
| **Northeast Brazil** | 2006-01 | 2006-03 | 2367 | 9511 |
| **South Brazil** | 2006-01 | 2006-05 | 639 | 2076 |
| States | | | | |
| **Bahia** | 2012-07 | 2012-06 | 867 | 1149 |
| **Ceará** | 2006-01 | 2009-07 | 1393 | 2307 |
| **Pernambuco** | 2008-07 | 2015-01 | 388 | 444 |
| **Piaui** | 2008-12 | 2015-06 | 535 | 646 |
| **Rio Grande do Norte** | 2006-01 | 2006-03 | 1393 | 2307 |
| **Rio Grande do Sul** | 2006-12 | 2006-05 | 1301 | 1961 |
| **Santa Catarina** | 2006-01 | 2014-01 | 255 | 288 |
| Single windparks | | | | |
| **Macaubas** | 2012-07 | 2012-06 | 78 | 78 |
| **Praia Formosa** | 2009-06 | 2009-07 | 78 | 78 |
| **Sao Clemente** | 2016-04 | 2016-04 | 87 | 87 |
| **Araripe** | 2016-11 | 2016-12 | 78 | 78 |
| **Alegria II** | 2011-12 | 2012-01 | 78 | 78 |
| **Elebras Cidreira 1** | 2011-07 | 2011-05 | 78 | 78 |
| **Bom Jardim** | 2011-10 | 2014-01 | 78 | 78 |

## A.2  Additional results
This part contains supplementary results of the statistical analysis of the simulated time series. On the one hand, tables (Table A 2, Table A 3, Table A 4) with the statistical parameters correlations, root mean square errors, mean bias errors (MBEs) and means of observed and simulated daily wind power generation are provided on different spatial levels (Brazil, subsystems, states, individual wind parks). On the other hand, boxplots (Figure A 1 - Figure A 7) of daily wind power generation are provided, for an easy comparison of the ranges of simulated and observed daily wind power generation. The results of the statistical analysis will not be discussed in detail, as this has been done in the main section.

*Table A 2: Correlations of observed and simulated daily wind power generation time series with three interpolation methods (Nearest Neighbour: NN, Bilinear Interpolation: BLI, Inverse Distance Weighting: IDW) for Brazil, its North-East and South, seven states and seven selected wind parks*

| | Correlation | | | RMSE [GWh] | | | MBE [GWh] | | | Mean [GWh] | | | |
| --- | --- | --- | --- | --- | --- | --- | --- | --- | --- | --- | --- | --- | --- |
| | NN | BLI | IDW | NN | BLI | IDW | NN | BLI | IDW | NN | BLI | IDW | obs |
| **Brazil** | 0.987 | 0.986 | 0.986 | 13.591 | 12.860 | 12.921 | 8.899 | 8.176 | 8.174 | 32.007 | 31.284 | 31.282 | 23.107 |
| **North-East** | 0.985 | 0.984 | 0.984 | 11.683 | 11.491 | 11.589 | 7.937 | 7.697 | 7.752 | 26.629 | 26.389 | 26.444 | 18.692 |
| **South** | 0.954 | 0.953 | 0.951 | 2.754 | 2.384 | 2.416 | 0.785 | 0.245 | 0.179 | 5.221 | 4.682 | 4.616 | 4.436 |
| **Bahia** | 0.955 | 0.952 | 0.950 | 2.541 | 2.639 | 2.726 | -0.014 | -0.236 | -0.365 | 9.167 | 8.945 | 8.815 | 9.180 |
| **Ceará** | 0.940 | 0.940 | 0.939 | 5.005 | 5.180 | 5.232 | 3.842 | 3.862 | 3.913 | 10.156 | 10.176 | 10.227 | 6.314 |
| **Pernambuco** | 0.876 | 0.875 | 0.873 | 1.546 | 1.543 | 1.550 | 0.411 | 0.379 | 0.362 | 4.864 | 4.832 | 4.816 | 4.454 |
| **Piaui** | 0.921 | 0.917 | 0.920 | 2.748 | 2.730 | 2.641 | 1.275 | 1.209 | 1.104 | 8.694 | 8.627 | 8.523 | 7.419 |
| **Rio Grande do Norte** | 0.974 | 0.973 | 0.973 | 5.911 | 5.735 | 5.879 | 3.987 | 3.859 | 3.951 | 11.593 | 11.465 | 11.557 | 7.606 |
| **Rio Grande do Sul** | 0.950 | 0.950 | 0.947 | 2.695 | 2.338 | 2.375 | 0.692 | 0.134 | 0.054 | 5.042 | 4.484 | 4.404 | 4.350 |
| **Santa Catarina** | 0.653 | 0.676 | 0.676 | 0.512 | 0.521 | 0.522 | -0.038 | -0.051 | -0.030 | 0.834 | 0.821 | 0.842 | 0.872 |
| **Macaubas** | 0.607 | 0.596 | 0.611 | 0.207 | 0.203 | 0.196 | -0.157 | -0.150 | -0.141 | 0.187 | 0.194 | 0.203 | 0.344 |
| **Praia Formosa** | 0.756 | 0.767 | 0.766 | 0.328 | 0.375 | 0.449 | 0.120 | 0.210 | 0.303 | 0.859 | 0.949 | 1.043 | 0.739 |
| **Sao Clemente** | 0.572 | 0.580 | 0.579 | 0.873 | 0.869 | 0.870 | 0.357 | 0.321 | 0.318 | 2.855 | 2.819 | 2.816 | 2.498 |
| **Araripe** | 0.853 | 0.852 | 0.853 | 2.102 | 1.973 | 1.893 | 1.671 | 1.505 | 1.404 | 4.640 | 4.473 | 4.372 | 2.968 |
| **Alegria II** | 0.553 | 0.554 | 0.551 | 0.636 | 0.605 | 0.626 | 0.503 | 0.472 | 0.492 | 1.174 | 1.142 | 1.163 | 0.671 |
| **Elebras Cidreira 1** | 0.823 | 0.819 | 0.809 | 0.254 | 0.263 | 0.273 | -0.056 | -0.103 | -0.116 | 0.528 | 0.481 | 0.468 | 0.584 |
| **Bom Jardim** | 0.460 | 0.526 | 0.531 | 0.218 | 0.221 | 0.222 | -0.105 | -0.123 | -0.125 | 0.120 | 0.101 | 0.099 | 0.224 |

Figure A 1 shows a comparison of differences between simulated and observed daily wind power generation for seven wind power plants in Brazil (a comparison on the level of states, subsystems or the whole country are not shown, as differences between the three interpolation methods are negligible for these realms). In general, the graphs indicate that the simulations fit observed wind power generation well, only slightly over- (Praia Formosa, Alegria II, Araripe) or underestimating (Macaubas) observed generation. The most notable differences between the interpolation methods are observed in Araripe, although they are not substantive. Similar to larger areas, the simulations for particular wind parks are close to observed wind

power generation and although some bias between simulations and observed wind power generation is notable, it is mostly small.

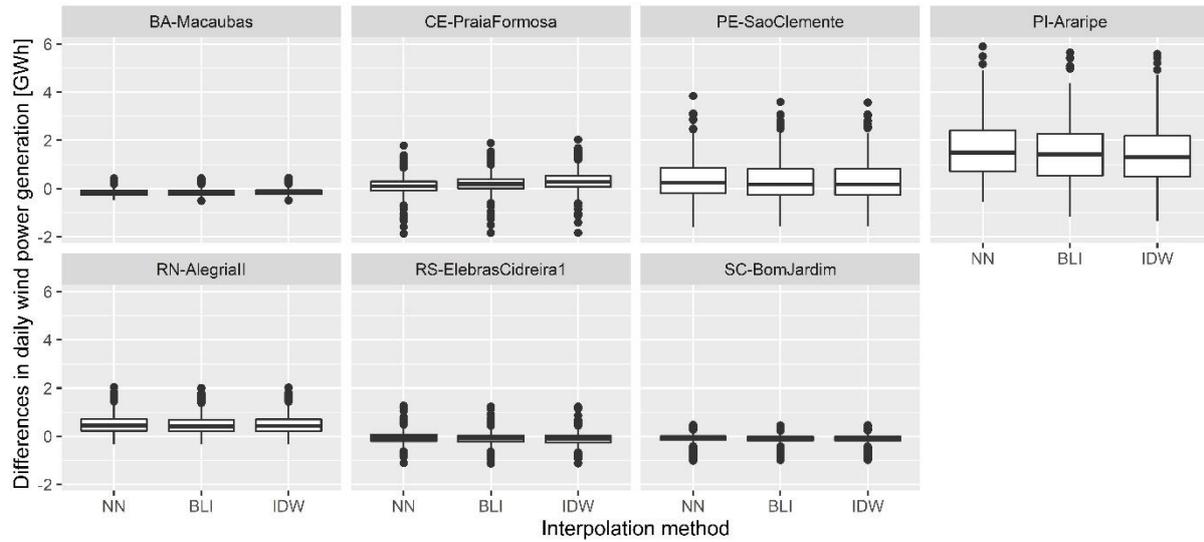

*Figure A 1: Comparison of differences between observed and simulated daily wind power generation with three interpolation methods (Nearest Neighbour: NN, Bilinear Interpolation: BLI, Inverse Distance Weighting: IDW) for seven wind power plants in Brazil*

*Table A 3: Comparison of root mean square errors (RMSEs), mean bias errors (MBEs) and means of simulated daily wind power generation with different sources for wind speed mean approximation: INMET wind speed measurements (INMET), Global Wind Atlas mean wind speeds (GWA) and no correction (Nearest Neighbour, NN)*

|  | RMSE [GWh] | | | MBE [GWh] | | | Mean [GWh] | | | |
| --- | --- | --- | --- | --- | --- | --- | --- | --- | --- | --- |
|  | NN | INMET | GWA | NN | INMET | GWA | NN | INMET | GWA | obs |
| Brazil | 13.591 | 5.246 | 7.693 | 8.899 | 0.598 | 3.939 | 32.007 | 23.706 | 27.046 | 23.107 |
| North-East | 11.683 | 5.079 | 7.058 | 7.937 | 0.426 | 3.705 | 26.629 | 19.118 | 22.398 | 18.692 |
| South | 2.754 | 2 | 2.101 | 0.785 | 0.114 | 0.059 | 5.221 | 4.55 | 4.495 | 4.436 |
| Bahia | 2.541 | 2.967 | 3.007 | -0.014 | -0.455 | -1.027 | 9.167 | 8.725 | 8.154 | 9.18 |
| Ceará | 5.005 | 1.789 | 4.162 | 3.842 | -0.57 | 3.143 | 10.156 | 5.744 | 9.456 | 6.314 |
| Pernambuco | 1.546 | 3.981 | 1.856 | 0.411 | -3.304 | -1.226 | 4.864 | 1.149 | 3.228 | 4.454 |
| Piaui | 2.748 | 2.23 | 2.36 | 1.275 | -0.834 | -0.183 | 8.694 | 6.585 | 7.235 | 7.419 |
| Rio Grande do Norte | 5.911 | 3.196 | 3.62 | 3.987 | 1.736 | 1.642 | 11.593 | 9.343 | 9.248 | 7.606 |
| Rio Grande do Sul | 2.695 | 2.007 | 2.027 | 0.692 | -0.346 | -0.105 | 5.042 | 4.003 | 4.245 | 4.35 |
| Santa Catarina | 0.512 | 0.877 | 0.514 | -0.038 | 0.493 | -0.014 | 0.834 | 1.365 | 0.858 | 0.872 |
| Macaubas | 0.207 | 0.207 | 0.193 | -0.157 | -0.157 | -0.138 | 0.187 | 0.187 | 0.206 | 0.344 |
| Praia Formosa | 0.328 | 0.328 | 0.342 | 0.12 | 0.12 | 0.149 | 0.859 | 0.859 | 0.888 | 0.739 |
| Sao Clemente | 0.873 | 2.218 | 0.898 | 0.357 | -2.06 | -0.481 | 2.855 | 0.438 | 2.017 | 2.498 |
| Araripe | 2.102 | 2.304 | 1.428 | 1.671 | -1.635 | 0.848 | 4.64 | 1.334 | 3.816 | 2.968 |
| Alegria II | 0.636 | 0.359 | 0.698 | 0.503 | -0.189 | 0.576 | 1.174 | 0.482 | 1.247 | 0.671 |
| Elebras Cidreira 1 | 0.254 | 0.348 | 0.26 | -0.056 | -0.24 | 0.042 | 0.528 | 0.344 | 0.626 | 0.584 |
| Bom Jardim | 0.218 | 0.337 | 0.224 | -0.105 | 0.05 | -0.116 | 0.12 | 0.274 | 0.108 | 0.224 |

The graphs in Figure A 2 to Figure A 4 show slightly different results than from the statistical analysis: Except for Macaubas and Praia Formosa where all simulations are in about the same range, the smallest differences between simulated and observed wind power generation are either the ones without wind speed mean approximation or when GWA data are applied. Only in Brazil, its North-East, Ceará and Alegria II the simulation with INMET mean wind speed approximation fits the range of observed daily wind power generation better than the other methods.

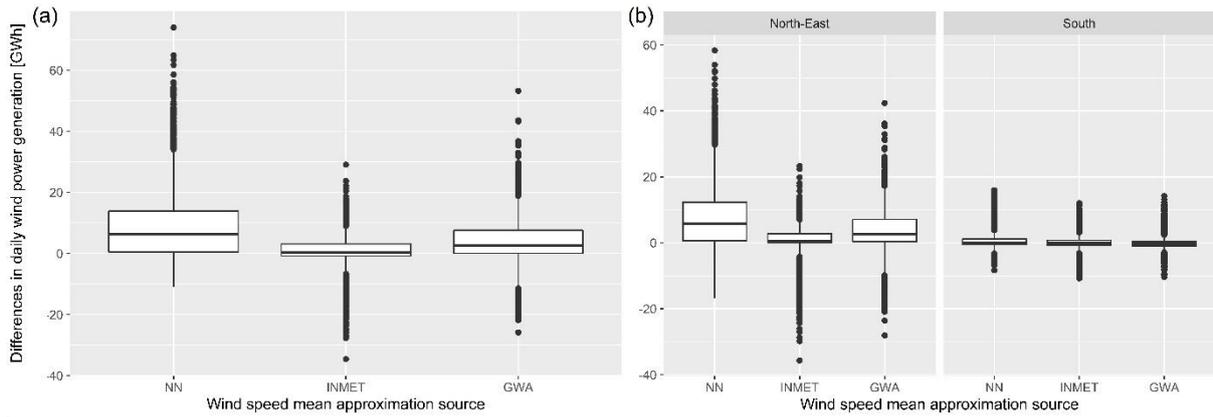

*Figure A 2: Comparison of differences between observed and simulated daily wind power generation with different wind speed mean approximation methods (no correction/Nearest Neighbour: NN, correction with measured wind speeds: INMET, Global Wind Atlas: GWA) for Brazil (a), its North-East and South (b)*

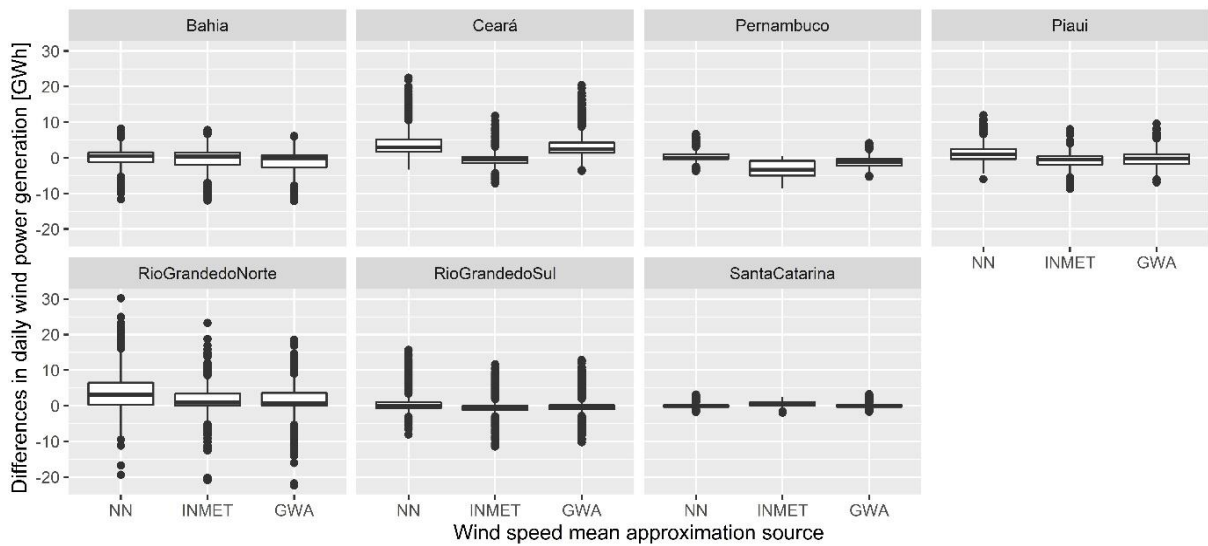

*Figure A 3: Comparison of differences between observed and simulated daily wind power generation with different wind speed mean approximation methods (no correction/Nearest Neighbour: NN, correction with measured wind speeds: INMET, Global Wind Atlas: GWA) for seven states of Brazil*

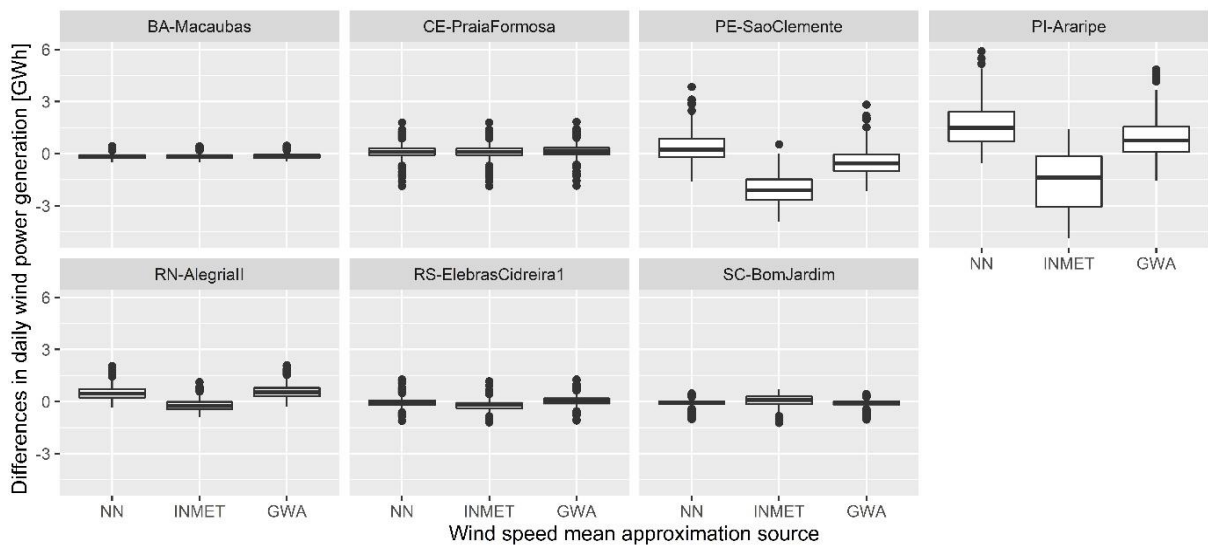

*Figure A 4: Comparison of differences between observed and simulated daily wind power generation with different wind speed mean approximation methods (no correction/Nearest Neighbour: NN, correction with measured wind speeds: INMET, Global Wind Atlas: GWA) for seven wind parks in Brazil*

Comparison of simulation results by boxplots reveals different findings than those of the statistical analysis for the spatiotemporal bias correction. In Figure A 5, for example, the graphs of daily wind power generation in Brazil, as well as in the North-East, indicate that wind power is simulated better if hourly and monthly wind speed correction is applied, whereas mean approximation leads to a higher overestimation of observed wind power generation. This partly contrasts the findings presented previously in Figure 9 and Table A 4, where sometimes lower RMSEs were obtained with mean approximation.

Similar to results from Brazil and the subsystems, on the level of single states (Figure A 6), the simulations with hourly and monthly wind speed correction usually are closer to observed wind power generation than with bias correction with GWA. These results are only partly supported by those from the statistical analysis, where the hourly and monthly correction often did not deliver a closer fit to observed wind power generation for most of the cases.

The graphs in Figure A 7 illustrate how wind speed bias correction affects wind power generation at particular locations. Of the two corrected wind parks Alegria II and Elebras Cidreira 1, the differences can be decreased and show a better fit in Alegria II while in Elebras Cidreira 1 a slightly larger error can be observed after correction. These results support the importance of choosing limitations for wind speed correction (distance to closest wind speed measurement station and correlation of wind speeds). With these limits, wind speed correction is applied on two wind parks (Elebras Cidreira 1 and Alegria II) with hourly and monthly wind speed correction.

*Table A 4: Comparison of correlations, root mean square errors (RMSEs), mean bias errors (MBEs) and means of simulated daily wind power generation with different methods for wind speed bias correction: mean approximation with Global Wind Atlas data (GWA) and mean approximation with Global Wind Atlas data combined with hourly and monthly wind speed correction with INMET wind speed data (GWA$_{hm}$)*

|  | Correlation | | RMSE [GWh] | | MBE [GWh] | | Mean [GWh] | | |
|---|---|---|---|---|---|---|---|---|---|
|  | GWA$_{hm}$ | GWA | GWA$_{hm}$ | GWA | GWA$_{hm}$ | GWA | GWA$_{hm}$ | GWA | obs |
| Brazil | 0.982 | 0.986 | 7.693 | 7.718 | 3.939 | -1.668 | 27.046 | 21.439 | 23.107 |
| North-East | 0.979 | 0.986 | 7.058 | 6.186 | 3.705 | -0.220 | 22.398 | 18.473 | 18.692 |
| South | 0.953 | 0.929 | 2.101 | 3.184 | 0.059 | -1.695 | 4.495 | 2.741 | 4.436 |
| Bahia | 0.954 | 0.950 | 3.007 | 4.457 | -1.027 | -2.238 | 8.154 | 6.942 | 9.180 |
| Ceará | 0.941 | 0.949 | 4.162 | 2.195 | 3.143 | 1.443 | 9.456 | 7.756 | 6.314 |
| Pernambuco | 0.856 | 0.873 | 1.856 | 2.063 | -1.226 | -1.575 | 3.228 | 2.879 | 4.454 |
| Piaui | 0.905 | 0.905 | 2.360 | 2.360 | -0.183 | -0.183 | 7.235 | 7.235 | 7.419 |
| Rio Grande do Norte | 0.965 | 0.975 | 3.620 | 3.283 | 1.642 | -0.321 | 9.248 | 7.285 | 7.606 |
| Rio Grande do Sul | 0.952 | 0.927 | 2.027 | 3.367 | -0.105 | -1.955 | 4.245 | 2.394 | 4.350 |
| Santa Catarina | 0.667 | 0.667 | 0.514 | 0.514 | -0.014 | -0.014 | 0.858 | 0.858 | 0.872 |
| Macaubas | 0.611 | 0.611 | 0.193 | 0.193 | -0.138 | -0.138 | 0.206 | 0.206 | 0.344 |
| Praia Formosa | 0.758 | 0.758 | 0.342 | 0.342 | 0.149 | 0.149 | 0.888 | 0.888 | 0.739 |
| Sao Clemente | 0.588 | 0.588 | 0.898 | 0.898 | -0.481 | -0.481 | 2.017 | 2.017 | 2.498 |
| Araripe | 0.882 | 0.882 | 1.428 | 1.428 | 0.848 | 0.848 | 3.816 | 3.816 | 2.968 |
| Alegria II | 0.555 | 0.554 | 0.698 | 0.447 | 0.576 | -0.335 | 1.247 | 0.336 | 0.671 |
| Elebras Cidreira 1 | 0.830 | 0.664 | 0.260 | 0.466 | 0.042 | -0.343 | 0.626 | 0.241 | 0.584 |
| Bom Jardim | 0.455 | 0.455 | 0.224 | 0.224 | -0.116 | -0.116 | 0.108 | 0.108 | 0.224 |

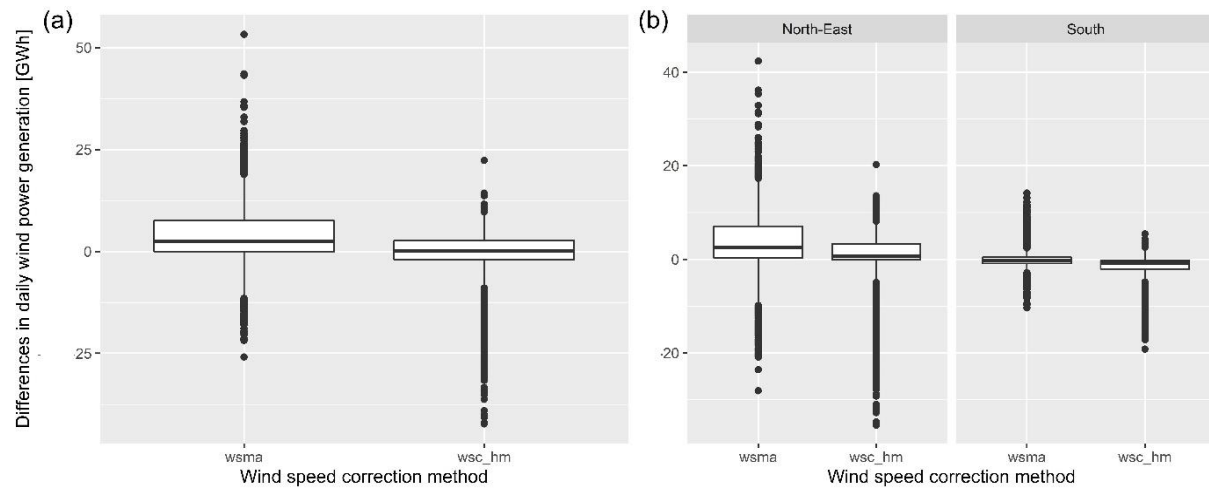

*Figure A 5: Comparison of differences between observed and simulated daily wind power generation with different wind speed bias correction methods (mean approximation with Global Wind Atlas wind speeds: wsma and mean approximation combined with hourly and monthly wind speed correction with INMET wind speeds: wsc_hm) for Brazil (a), its North-East and South (b)*

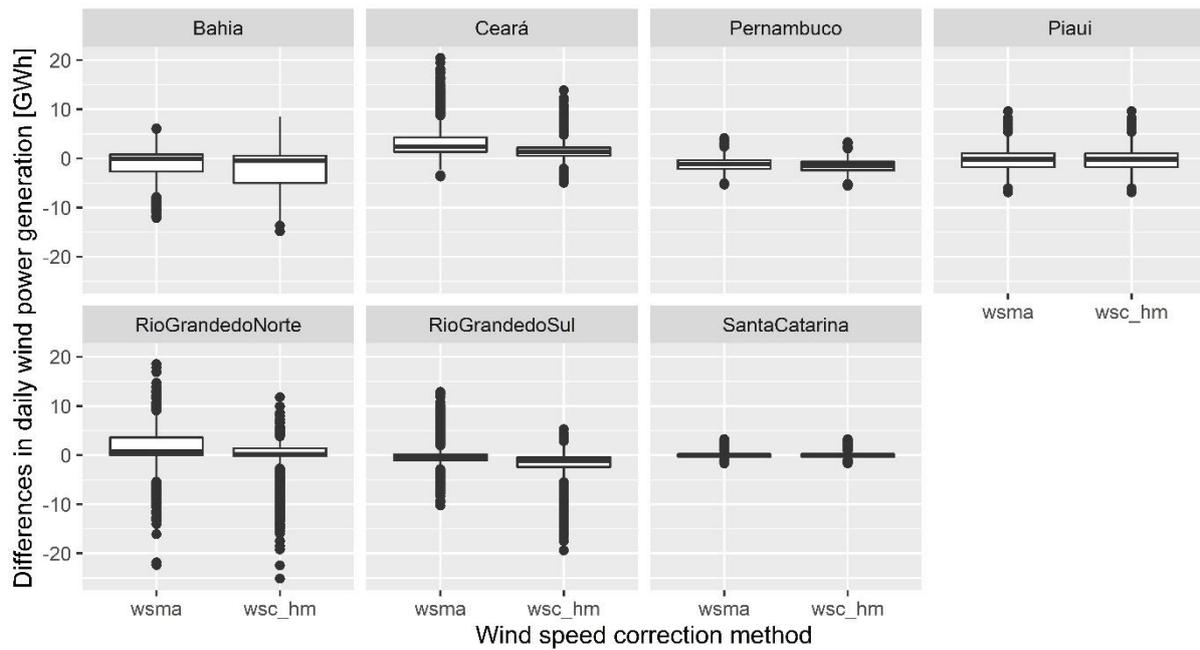

*Figure A 6: Comparison of differences between observed and simulated daily wind power generation with different wind speed bias correction methods (mean approximation with Global Wind Atlas wind speeds: wsma and mean approximation combined with hourly and monthly wind speed correction with INMET wind speeds: wsc_hm) for Brazilian states*

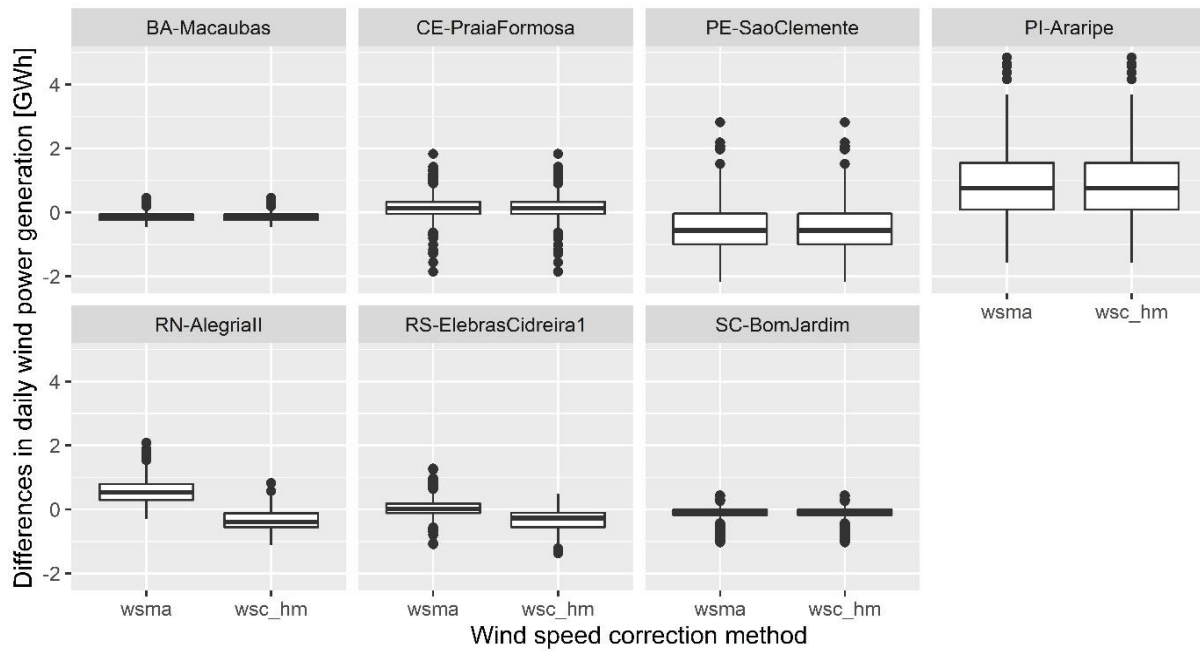

*Figure A 7: Comparison of differences between observed and simulated daily wind power generation with different wind speed bias correction methods (mean approximation with Global Wind Atlas wind speeds: wsma and mean approximation combined with hourly and monthly wind speed correction with INMET wind speeds: wsc_hm) for seven Brazilian wind power plants*

## A.3 Results from other studies

Table A5 shows results from other studies to compare the values from the present study to. A description can be found in the Discussion.

*Table A 5: Collection of statistical parameters (correlations, rel. RMSEs and rel. MBEs) from other studies for comparison to results from the present study. The results of González-Aparicio et al. [1] were given as absolute values, relative values were calculated from those by normalising by the installed capacity*

| Source | Dataset | Region | Temporal resolution | Correlation | Rel. RMSE | Rel. Bias | Notes |
|---|---|---|---|---|---|---|---|
| **Cannon et al. [14]** | MERRA | Great Britain | hourly | 0.96 | | | |
| **Cradden et al. [17]** | MERRA | Ireland | monthly | 0.97 | 10.17% | -0.79% | |
| **Pfenninger and Staffell [5]** | MERRA | Germany | hourly | 0.981 | 3.11% | | Correlation calculated from $R^2$ |
| **Pfenninger and Staffell [5]** | MERRA | Spain | hourly | 0.917 | 6.07% | | Correlation calculated from $R^2$ |
| **Pfenninger and Staffell [5]** | MERRA | Britain | hourly | 0.967 | 4.68% | | Correlation calculated from $R^2$ |
| **Pfenninger and Staffell [5]** | MERRA | France | hourly | 0.956 | 4.39% | | Correlation calculated from $R^2$ |
| **Pfenninger and Staffell [5]** | MERRA | Italy | hourly | 0.872 | 7.44% | | Correlation calculated from $R^2$ |
| **Pfenninger and Staffell [5]** | MERRA | Sweden | hourly | 0.952 | 5.66% | | Correlation calculated from $R^2$ |
| **Pfenninger and Staffell [5]** | MERRA | Denmark | hourly | 0.955 | 6.75% | | Correlation calculated from $R^2$ |
| **Pfenninger and Staffell [5]** | MERRA | Ireland | hourly | 0.951 | 6.65% | | Correlation calculated from $R^2$ |
| **Pfenninger and Staffell [5]** | MERRA | EU (13 European countries) | monthly | 0.951 | | | Mean correlation for 13 countries, correlation calculated from $R^2$ |
| **Pfenninger and Staffell [5]** | MERRA | Germany | monthly | 0.991 | | | Correlation calculated from $R^2$ |
| **Pfenninger and Staffell [5]** | MERRA | Ireland | monthly | 0.986 | | | Correlation calculated from $R^2$ |
| **Pfenninger and Staffell [5]** | MERRA | France | monthly | 0.983 | | | Correlation calculated from $R^2$ |
| **Pfenninger and Staffell [5]** | MERRA | Denmark | monthly | 0.977 | | | Correlation calculated from $R^2$ |
| **Pfenninger and Staffell [5]** | MERRA | Spain | monthly | 0.975 | | | Correlation calculated from $R^2$ |
| **Pfenninger and Staffell [5]** | MERRA | Great Britain | monthly | 0.969 | | | Correlation calculated from $R^2$ |
| **Pfenninger and Staffell [5]** | MERRA | Norway | monthly | 0.967 | | | Correlation calculated from $R^2$ |
| **Pfenninger and Staffell [5]** | MERRA | Sweden | monthly | 0.957 | | | Correlation calculated from $R^2$ |
| **Pfenninger and Staffell [5]** | MERRA | Finland | monthly | 0.952 | | | Correlation calculated from $R^2$ |
| **Pfenninger and Staffell [5]** | MERRA | Romania | monthly | 0.933 | | | Correlation calculated from $R^2$ |
| **Pfenninger and Staffell [5]** | MERRA | Portugal | monthly | 0.922 | | | Correlation calculated from $R^2$ |
| **Pfenninger and Staffell [5]** | MERRA | Italy | monthly | 0.920 | | | Correlation calculated from $R^2$ |
| **Pfenninger and Staffell [5]** | MERRA | Greece | monthly | 0.873 | | | Correlation calculated from $R^2$ |
| **Pfenninger and Staffell [5]** | MERRA-2 | EU (13 European countries) | monthly | 0.955 | | | mean correlation for 13 countries, correlation calculated from $R^2$ |
| **Pfenninger and Staffell [5]** | MERRA-2 | Germany | monthly | 0.989 | | | Correlation calculated from $R^2$ |
| **Pfenninger and Staffell [5]** | MERRA-2 | Ireland | monthly | 0.985 | | | Correlation calculated from $R^2$ |
| **Pfenninger and Staffell [5]** | MERRA-2 | France | monthly | 0.982 | | | Correlation calculated from $R^2$ |
| **Pfenninger and Staffell [5]** | MERRA-2 | Denmark | monthly | 0.973 | | | Correlation calculated from $R^2$ |
| **Pfenninger and Staffell [5]** | MERRA-2 | Spain | monthly | 0.976 | | | Correlation calculated from $R^2$ |
| **Pfenninger and Staffell [5]** | MERRA-2 | Great Britain | monthly | 0.969 | | | Correlation calculated from $R^2$ |
| **Pfenninger and Staffell [5]** | MERRA-2 | Norway | monthly | 0.968 | | | Correlation calculated from $R^2$ |
| **Pfenninger and Staffell [5]** | MERRA-2 | Sweden | monthly | 0.947 | | | Correlation calculated from $R^2$ |
| **Pfenninger and Staffell [5]** | MERRA-2 | Finland | monthly | 0.955 | | | Correlation calculated from $R^2$ |
| **Pfenninger and Staffell [5]** | MERRA-2 | Romania | monthly | 0.935 | | | Correlation calculated from $R^2$ |
| **Pfenninger and Staffell [5]** | MERRA-2 | Portugal | monthly | 0.875 | | | Correlation calculated from $R^2$ |
| **Pfenninger and Staffell [5]** | MERRA-2 | Italy | monthly | 0.925 | | | Correlation calculated from $R^2$ |
| **Pfenninger and Staffell [5]** | MERRA-2 | Greece | monthly | 0.885 | | | Correlation calculated from $R^2$ |

| Reference | Dataset | Country | Resolution | Correlation | Error | Bias | Notes |
|---|---|---|---|---|---|---|---|
| **González-Aparicio et al. [39]** | MERRA | Austria | hourly | 0.904 | 12.9% | 0.6% | only 2015 |
| **González-Aparicio et al. [39]** | MERRA | Belgium | hourly | 0.937 | 4.5% | -1.4% | only 2015 |
| **González-Aparicio et al. [39]** | MERRA | Bulgaria | hourly | 0.759 | 22.2% | -15.5% | only 2015 |
| **González-Aparicio et al. [39]** | MERRA | Cyprus | hourly | 0.436 | 13.2% | -6.9% | only 2015 |
| **González-Aparicio et al. [39]** | MERRA | Czech Republic | hourly | 0.92 | 11.8% | 1.0% | only 2015 |
| **González-Aparicio et al. [39]** | MERRA | Germany | hourly | 0.971 | 3.8% | 0.7% | only 2015 |
| **González-Aparicio et al. [39]** | MERRA | Denmark | hourly | 0.952 | 4.9% | 0.0% | only 2015 |
| **González-Aparicio et al. [39]** | MERRA | Estonia | hourly | 0.913 | 7.8% | 0.5% | only 2015 |
| **González-Aparicio et al. [39]** | MERRA | Spain | hourly | 0.916 | 9.4% | 1.6% | only 2015 |
| **González-Aparicio et al. [39]** | MERRA | Finland | hourly | 0.929 | 9.1% | 1.9% | only 2015 |
| **González-Aparicio et al. [39]** | MERRA | France | hourly | 0.952 | 5.9% | 1.4% | only 2015 |
| **González-Aparicio et al. [39]** | MERRA | Greece | hourly | 0.816 | 11.4% | -1.8% | only 2015 |
| **González-Aparicio et al. [39]** | MERRA | Croatia | hourly | 0.788 | 15.7% | -6.5% | only 2015 |
| **González-Aparicio et al. [39]** | MERRA | Hungary | hourly | 0.897 | 12.2% | -6.0% | only 2015 |
| **González-Aparicio et al. [39]** | MERRA | Ireland | hourly | 0.964 | 6.5% | 0.6% | only 2015 |
| **González-Aparicio et al. [39]** | MERRA | Lithuania | hourly | 0.923 | 12.0% | -5.4% | only 2015 |
| **González-Aparicio et al. [39]** | MERRA | Latvia | hourly | 0.905 | 7.0% | 0.0% | only 2015 |
| **González-Aparicio et al. [39]** | MERRA | Netherlands | hourly | 0.949 | 12.9% | -11.3% | only 2015 |
| **González-Aparicio et al. [39]** | MERRA | Poland | hourly | 0.967 | 5.0% | 0.1% | only 2015 |
| **González-Aparicio et al. [39]** | MERRA | Portugal | hourly | 0.824 | 13.0% | -4.9% | only 2015 |
| **González-Aparicio et al. [39]** | MERRA | Romania | hourly | 0.838 | 14.1% | -6.3% | only 2015 |
| **González-Aparicio et al. [39]** | MERRA | Sweden | hourly | 0.866 | 30.7% | 5.4% | only 2015 |
| **González-Aparicio et al. [39]** | MERRA | United Kingdom | hourly | 0.863 | 4.4% | -0.9% | only 2015 |
| **González-Aparicio et al. [39]** | MERRA | Switzerland | hourly | 0.581 | 20.5% | 8.5% | only 2015 |
| **González-Aparicio et al. [39]** | EMHIRES | Austria | hourly | 0.869 | 14.0% | -1.2% | with GWA bias correction |
| **González-Aparicio et al. [39]** | EMHIRES | Belgium | hourly | 0.947 | 4.2% | -0.2% | with GWA bias correction |
| **González-Aparicio et al. [39]** | EMHIRES | Bulgaria | hourly | 0.733 | 22.3% | -15.6% | with GWA bias correction |
| **González-Aparicio et al. [39]** | EMHIRES | Cyprus | hourly | 0.427 | 13.9% | -0.8% | with GWA bias correction |
| **González-Aparicio et al. [39]** | EMHIRES | Czech Republic | hourly | 0.904 | 17.2% | 3.4% | with GWA bias correction |
| **González-Aparicio et al. [39]** | EMHIRES | Germany | hourly | 0.972 | 7.3% | 2.9% | with GWA bias correction |
| **González-Aparicio et al. [39]** | EMHIRES | Denmark | hourly | 0.957 | 5.4% | 1.4% | with GWA bias correction |
| **González-Aparicio et al. [39]** | EMHIRES | Estonia | hourly | 0.920 | 8.1% | 1.3% | with GWA bias correction |
| **González-Aparicio et al. [39]** | EMHIRES | Spain | hourly | 0.913 | 9.8% | 1.7% | with GWA bias correction |
| **González-Aparicio et al. [39]** | EMHIRES | Finland | hourly | 0.944 | 6.0% | 0.6% | with GWA bias correction |
| **González-Aparicio et al. [39]** | EMHIRES | France | hourly | 0.959 | 6.3% | 1.7% | with GWA bias correction |
| **González-Aparicio et al. [39]** | EMHIRES | Greece | hourly | 0.813 | 11.5% | -1.7% | with GWA bias correction |
| **González-Aparicio et al. [39]** | EMHIRES | Croatia | hourly | 0.814 | 14.5% | -6.5% | with GWA bias correction |
| **González-Aparicio et al. [39]** | EMHIRES | Hungary | hourly | 0.876 | 12.9% | -6.0% | with GWA bias correction |
| **González-Aparicio et al. [39]** | EMHIRES | Ireland | hourly | 0.965 | 6.6% | 0.9% | with GWA bias correction |
| **González-Aparicio et al. [39]** | EMHIRES | Lithuania | hourly | 0.926 | 11.4% | -5.4% | with GWA bias correction |
| **González-Aparicio et al. [39]** | EMHIRES | Latvia | hourly | 0.921 | 6.7% | 0.0% | with GWA bias correction |
| **González-Aparicio et al. [39]** | EMHIRES | Netherlands | hourly | 0.960 | 12.3% | -9.2% | with GWA bias correction |
| **González-Aparicio et al. [39]** | EMHIRES | Poland | hourly | 0.965 | 7.2% | 1.8% | with GWA bias correction |
| **González-Aparicio et al. [39]** | EMHIRES | Portugal | hourly | 0.846 | 12.6% | -2.8% | with GWA bias correction |
| **González-Aparicio et al. [39]** | EMHIRES | Romania | hourly | 0.836 | 14.1% | -6.3% | with GWA bias correction |
| **González-Aparicio et al. [39]** | EMHIRES | Sweden | hourly | 0.885 | 28.8% | 6.1% | with GWA bias correction |
| **González-Aparicio et al. [39]** | EMHIRES | United Kingdom | hourly | 0.855 | 4.2% | -0.5% | with GWA bias correction |

| Reference | Dataset | Country | Resolution | Correlation | MAE | Bias | Notes |
|---|---|---|---|---|---|---|---|
| **González-Aparicio et al. [39]** | EMHIRES | Switzerland | hourly | 0.545 | 21.7% | 8.5% | with GWA bias correction |
| **González-Aparicio et al. [39]** | ECMWF | Austria | hourly | 0.904 | 9.8% | 1.0% | non-freely available dataset |
| **González-Aparicio et al. [39]** | ECMWF | Belgium | hourly | 0.937 | 3.0% | 0.0% | non-freely available dataset |
| **González-Aparicio et al. [39]** | ECMWF | Bulgaria | hourly | 0.759 | 12.7% | -3.5% | non-freely available dataset |
| **González-Aparicio et al. [39]** | ECMWF | Cyprus | hourly | 0.436 | 12.6% | -3.7% | non-freely available dataset |
| **González-Aparicio et al. [39]** | ECMWF | Czech Republic | hourly | 0.920 | 17.5% | 2.1% | non-freely available dataset |
| **González-Aparicio et al. [39]** | ECMWF | Germany | hourly | 0.971 | 4.4% | 1.6% | non-freely available dataset |
| **González-Aparicio et al. [39]** | ECMWF | Denmark | hourly | 0.952 | 4.2% | -0.2% | non-freely available dataset |
| **González-Aparicio et al. [39]** | ECMWF | Estonia | hourly | 0.913 | 7.0% | 1.1% | non-freely available dataset |
| **González-Aparicio et al. [39]** | ECMWF | Spain | hourly | 0.916 | 7.2% | 0.7% | non-freely available dataset |
| **González-Aparicio et al. [39]** | ECMWF | Finland | hourly | 0.929 | 6.5% | 1.1% | non-freely available dataset |
| **González-Aparicio et al. [39]** | ECMWF | France | hourly | 0.952 | 5.9% | 1.9% | non-freely available dataset |
| **González-Aparicio et al. [39]** | ECMWF | Greece | hourly | 0.816 | 11.0% | -2.2% | non-freely available dataset |
| **González-Aparicio et al. [39]** | ECMWF | Croatia | hourly | 0.788 | 18.2% | -6.5% | non-freely available dataset |
| **González-Aparicio et al. [39]** | ECMWF | Hungary | hourly | 0.897 | 10.8% | -5.9% | non-freely available dataset |
| **González-Aparicio et al. [39]** | ECMWF | Ireland | hourly | 0.964 | 11.6% | 0.0% | non-freely available dataset |
| **González-Aparicio et al. [39]** | ECMWF | Lithuania | hourly | 0.923 | 10.7% | -5.5% | non-freely available dataset |
| **González-Aparicio et al. [39]** | ECMWF | Latvia | hourly | 0.905 | 5.9% | 0.0% | non-freely available dataset |
| **González-Aparicio et al. [39]** | ECMWF | Netherlands | hourly | 0.949 | 6.5% | 2.0% | non-freely available dataset |
| **González-Aparicio et al. [39]** | ECMWF | Poland | hourly | 0.967 | 4.7% | 1.0% | non-freely available dataset |
| **González-Aparicio et al. [39]** | ECMWF | Portugal | hourly | 0.824 | 14.2% | -3.9% | non-freely available dataset |
| **González-Aparicio et al. [39]** | ECMWF | Romania | hourly | 0.838 | 12.8% | -6.4% | non-freely available dataset |
| **González-Aparicio et al. [39]** | ECMWF | Sweden | hourly | 0.866 | 30.3% | 6.4% | non-freely available dataset |
| **González-Aparicio et al. [39]** | ECMWF | Slovenia | hourly | | 26.7% | 16.7% | non-freely available dataset |
| **González-Aparicio et al. [39]** | ECMWF | United Kingdom | hourly | 0.863 | 3.7% | -1.8% | non-freely available dataset |
| **González-Aparicio et al. [39]** | ECMWF | Switzerland | hourly | 0.581 | 34.0% | 8.5% | non-freely available dataset |

**A.4 Extrapolation to hub height**

Using the power law with wind speeds at 10 m and 50 m height may not be an appropriate method for extrapolation of wind speeds to hub height and would result in inaccurate wind speeds, especially in complex terrain. The authors therefore tested another method, the log wind profile with logarithmic least squares fit, using wind speeds in three heights (2 m and 10 m above disposition height and 50 m above ground). According to Equation A1, *a* and *b* are fitted with regression to wind speeds $w^h$ at heights *h*, and then the desired height is inserted.

$$w^h = a + b * log(h) \tag{A1}$$

Furthermore, two other approaches are applied to the power law: Calculating the shear exponent α from the wind speeds at 2 m and 10 m above disposition height and (1) extrapolation from 10 m wind speeds or (2) extrapolation from 50 m wind speeds.

At first glance the logarithmic wind profile method looked promising: Figure A 8 shows, that errors can be decreased on all spatial levels by applying the log wind profile method, and correlations are increased on the level of wind parks.

However, two major disadvantages of this method should be mentioned: (1) This method requires more computational power and data storage and (2) in some cases it results in negative wind speeds at hub height, if the wind speed at 50 m height is lower than at the heights below. (2) has become a more serious problem, when addressing wind speeds at lower heights (10 m) for comparison to measured wind speeds during bias correction. In this case, 12 of 444 wind park locations resulted in negative wind speeds, ten of which had more than 40 % of negative wind speeds, five with more than 50 % of negative wind speeds, and three even over 98 % of negative wind speeds. Therefore, this method was not considered useful for this purpose, and it was decided to maintain the original power law method for extrapolation.

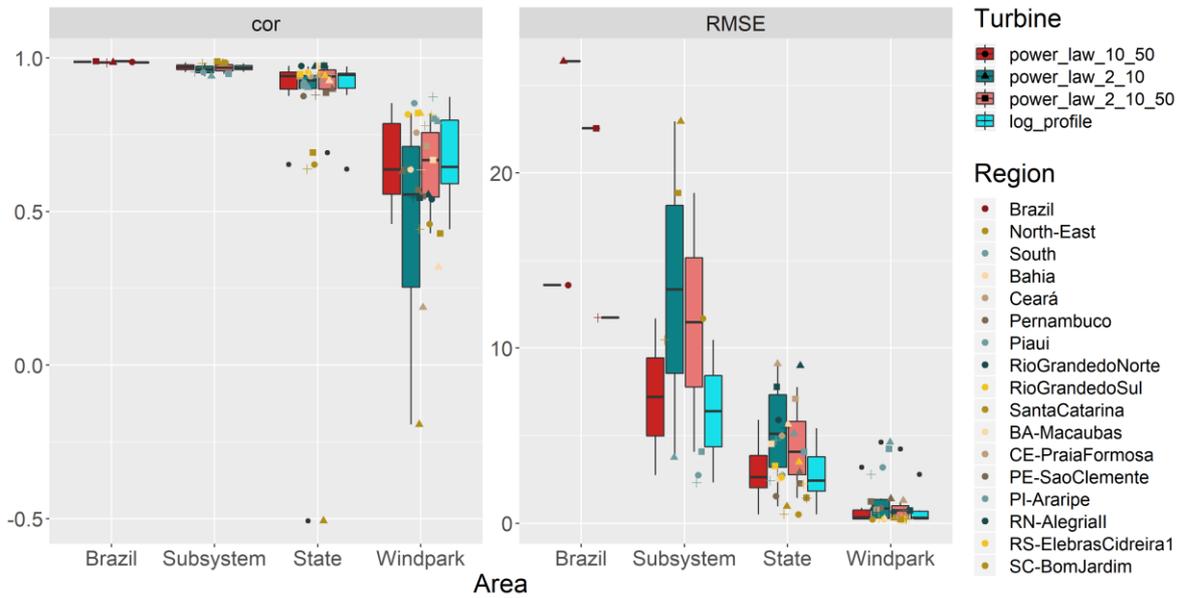

*Figure A 8: Comparison of power law and log wind profile with logarithmic least squares for extrapolation to hub height using Nearest Neighbour method. Power law either uses two heights: 2 m and 10 m above displacement height (power_law_2_10) or 10 m above displacement height and 50 m above ground (power_law_10_50) for calculation of shear exponent α, or three heights, of which the two lower were used for the calculation of α and 50 m wind speeds were applied for extrapolation (power_law_2_10_50)*

This method was also tested extracting wind speeds from different heights. Using wind speeds at lower heights (2 m and 10 m above disposition height) decreases correlations and increases the errors and is therefore not useful for this application. Also extrapolating from wind speeds in 50 m height while still using the wind shear exponent obtained from the wind speeds at lower heights, shows lower RMSEs and higher correlations compared to using only wind speeds from 2 m and 10 m height, but the results are still better when applying the power law to wind speeds in 10 m and 50 m height. The authors therefore decided to maintain the original method.

**A.5 Distance to wind speed measurement stations**

The maximum distance of a wind speed measurement station to the location of wind parks had to be determined as a measure of quality control. For that purpose, different distances were tested. In Figure A 9, simulation results for distances between 30 and 80 km are compared. Correlations hardly change when varying the maximum allowed distance, but errors (absolute MBEs and RMSEs) change for regional aggregations above the level of wind parks, where only a restriction of 30 km showed an impact. In most cases, a maximum distance of 40 km to the closest INMET wind speed measurement station leads to the smallest errors, which is consequently used in the simulation.

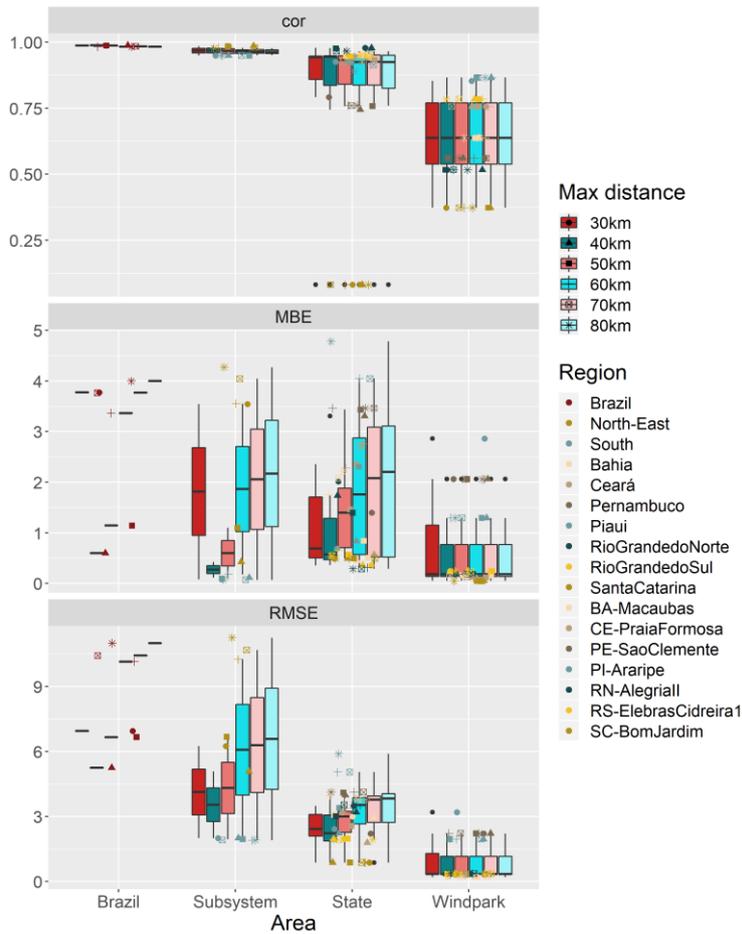

*Figure A 9: Comparison of different maximum allowed distances to INMET wind speed measurement stations for bias correction, using the Nearest Neighbour method and mean bias correction*

However, using a smaller limit for the distances to the wind speed measurement stations also comes with drawbacks: This allows less stations to be used for bias correction (see Figure A 10).

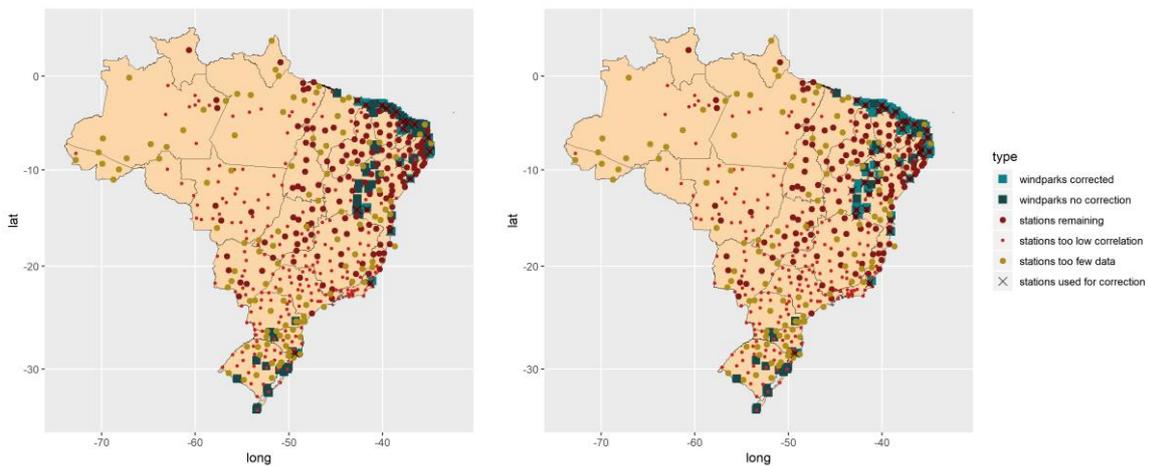

*Figure A 10: Comparison of corrected wind parks using 40 km (left) and 80 km (right) maximum allowed distances to INMET wind speed measurement stations for bias correction*

27